# Reinforcement Learning of Control Strategies for Reducing Skin Friction Drag in a Fully Developed Channel Flow

Takahiro Sonoda, Zhuchen Liu, Toshitaka Itoh, and Yosuke Hasegawa†

Institute of Industrial Science, The University of Tokyo
E-mail: ysk@iis.u-tokyo.ac.jp

**Abstract:**
Reinforcement learning is applied to the development of control strategies in order to reduce skin friction drag in a fully developed turbulent channel flow at a low Reynolds number. Motivated by the so-called opposition control (Choi *et al.* 1993), in which a control input is applied so as to cancel the wall-normal velocity fluctuation on a detection plane at a certain distance from the wall, we consider wall blowing and suction as a control input, and its spatial distribution is determined by the instantaneous streamwise and wall-normal velocity fluctuations at the distance of 15 wall units above the wall. Deep neural network is used to express the complex relationship between the sensing information and the control input, and it is trained so as to maximize the expected long-term reward, i.e., drag reduction. When only the wall-normal velocity fluctuation is measured and a linear network is used, the present framework successfully reproduces the optimal linear weight for the opposition control reported in the previous study (Chung & Talha 2011). In contrast, when a non-linear network is used, more complex control strategies based on the instantaneous streamwise and wall-normal velocity fluctuations are obtained. Specifically, the obtained control strategies abruptly switch between strong wall blowing and suction for downwelling of a high-speed fluid toward the wall and upwelling of a low-speed fluid away from the wall, respectively. The obtained control policies lead to drag reduction rates as high as 37 %, which is higher than 23 % achieved by the conventional opposition control at the same Reynolds number. Finding such an effective and non-linear control policy is quite difficult by solely relying on human insights. The present results indicate that reinforcement learning can be a novel framework for the development of effective control strategies through systematic learning based on a large number of trials.

## 1 Introduction

Turbulent flows are ubiquitous in our daily life and determine the performances and the energy efficiencies of various thermo-fluids devices (Brunton & Noack 2015). In most engineering flows, turbulence is bounded by a solid surface, and their interaction plays crucial roles in generation and maintenance of near-wall turbulence, and associated momentum and heat transport between fluid and solid. Even over simple geometries such as a smooth flat wall, however, turbulence exhibits complex behavior due to its non-linear and multiscale nature, so that prediction and control of turbulent flow remain challenging.

In this study, we consider the control of a fully developed turbulent channel flow, which is one of canonical flow configurations. Since near-wall turbulence is responsible for the increase in wall skin friction, a tremendous amount of efforts has been devoted to reducing the skin friction drag. In general, flow control strategies can be categorized into passive and active schemes (Gad-el Hak 1996). The passive scheme does not require power input for control, and its typical example is a riblet surface (Dean & Bhushan 2010). In contrast, the active scheme requires additional power input for control, and it can be further classified into predetermined and feedback controls. The former applies a control input with a predetermined spatio-temporal distribution regardless of an instantaneous flow state. This makes a control system simple since no sensing of a flow field is required. Despite its simplicity, it is known that the predetermined control achieves relatively high drag reduction rates, and various control modes such as spanwise wall oscillation (Jung *et al.* 1992; Quadrio & Ricco 2004), streamwise traveling wave of wall blowing and suction (Min *et al.* 2006; Lieu *et al.* 2010; Mamori *et al.* 2014), and uniform wall blowing (Sumitani & Kasagi 1995; Kametani & Fukagata 2011) have been proposed.

In contrast, the feedback control determines a control input based on a sensor signal obtained from an instantaneous flow field, and therefore it enables a more flexible control. Meanwhile, due to the large degrees of freedom of the flow state and also the control input, optimizing a feedback control law is quite challenging. Therefore, existing control strategies have often been developed based on researchers' physical insights. A typical example of a feedback control is the so-called opposition control (Choi *et al.* 1994; Hammond *et al.* 1998; Chung & Talha 2011), where local wall blowing and suction is applied so as to cancel the wall-normal velocity fluctuation at a certain height from the wall. The sensing plane is called a detection plane, and its optimal height has been reported as $y^+ = 15$ in a wall unit (Hammond *et al.* 1998). The relationship between the wall-normal velocity on the detection plane and the control input has commonly been assumed to be linear a priori, and its optimal weight coefficient was found to be around unity (Choi *et al.* 1994; Chung & Talha 2011). It should be noted that optimization of these parameters in the control algorithm have mostly been done through trials and errors, and such an approach is quite inefficient even for a simple control algorithm where the relationship between the sensor signal and the control input is assumed to be linear.

There also exists another approach to develop efficient feedback control laws. Optimal control theory is a powerful tool to optimize a control input with large degrees of freedom by explicitly leveraging mathematical models of a flow system such as Navier- Stokes equations and mass conservation. Specifically, the spatio-temporal distribution of a control input is determined so as to minimize a prescribed cost functional. The cost functional can be defined within a certain time horizon, so that the future flow dynamics are taken into consideration in the optimization procedures. Optimal control theory was successfully applied to a low-Reynolds-number turbulent channel flow by Bewley *et al.* (2001), and it was demonstrated that the flow can be relaminarized. One of the major drawbacks in optimal control theory is that it requires expensive iterations of forward and adjoint simulations within the time horizon in order to determine the optimal control input. By assuming a vanishingly small time horizon, suboptimal control theory (Lee *et al.* 1998; Hasegawa & Kasagi 2011) provides an analytical expression of the control input without solving adjoint equations, but its control performance is not as high as that achieved by optimal control theory, suggesting the importance of considering future flow dynamics in determining the control input. Another issue is that there exists sever limitation in the length of the time horizon employed in optimal control theory due to inherent instability of adjoint equations (Wang *et al.* 2014). Specifically, the maximum time horizon is around 100 in a wall unit (Bewley *et al.* 2001; Yamamoto *et al.* 2013), which is quite short considering the time-scale of wall turbulence. Especially, this limitation becomes critical at higher Reynolds numbers where large-scale structures play important roles in the dynamics of wall turbulence (Kim & Adrian 1999).

In recent years, much attention has been paid to reinforcement learning as a new framework for developing efficient control strategies in various fields such as robot control (Kober *et al.* 2013), games (Silver *et al.* 2016), to name a few. In reinforcement learning, an agent decides its action based on a current state. As a consequence, the agent receives a reward from an environment. By repeating this interaction with the environment, the agent learns an efficient policy, which dictates the relationship between the state and the action, so as to maximize the total expected future reward. In this way, the policy can be optimized from a long-term perspective. In addition, by combining reinforcement learning and deep neural networks, deep reinforcement learning (Sutton & Barto 2018) can naturally deal with a complex non-linear relationship between sensor signals and a control input.

Recent applications of reinforcement learning to fluid problems cover a variety of purposes such as drag reduction (Koizumi *et al.* 2018; Rabault *et al.* 2019; Rabault & Kuhnle 2019; Tokarev *et al.* 2020; Xu *et al.* 2020; Tang *et al.* 2020; Paris *et al.* 2021; Ghraieb *et al.* 2021; Ren *et al.* 2021; Fan *et al.* 2020), control of heat transfer (Beintema *et al.* 2020; Hachem *et al.* 2021), optimization of microfluidics (Lee *et al.* 2021; Dressler *et al.* 2018), optimization of artificial swimmers (Novati *et al.* 2018; Verma *et al.* 2018; Zhu *et al.* 2021; Yan *et al.* 2020) and shape optimization (Yan *et al.* 2019; Li *et al.* 2021; Viquerat *et al.* 2021; Qin *et al.* 2021). In terms of drag reduction considered in the present study, Rabault *et al.* (2019) considered control of a two-dimensional flow round a cylinder at a low Reynolds number. They assumed wall blowing and suction from two local slits over the cylinder, and demonstrated that a control policy obtained by reinforcement learning achieves 8% drag reduction. Tang *et al.* (2020) discussed Reynolds number effects on the control performance at different Reynolds numbers of 100, 200, 300 and 400, and also showed the possibility of applying the obtained control policy to unseen Reynolds numbers. Paris *et al.* (2021) optimized the arrangement of sensors employed for controlling 2D laminar flow behind a cylinder. Their sparsity-seeking algorithm allows to reduce a number of sensors down to five without sacrificing the control performance. Ghraieb *et al.* (2021) proposed a degenerated version of reinforcement learning so that it does not require the information of the state as an input. This allows to find effective open-loop control policies for both laminar and turbulent flows around an airfoil and a cylinder. Fan *et al.* (2020) experimentally demonstrated that reinforcement learning can find effective rotation modes of small cylinders around a primal stationary cylinder for its drag reduction. As shown above, most previous studies consider relatively simple flow fields such as a two-dimensional laminar flow around a blunt object. Also, their control inputs are wall blowing and suction from slots at two or four prescribed locations, or rotation/vibrations of one or two cylinders, so that the degrees of freedom for a control input are commonly limited. Therefore, it remains an open question whether reinforcement learning can be applicable to turbulence control with a control input having large degrees of freedom.

To the best of the authors' knowledge, this is the first study applying reinforcement learning to control of a fully developed turbulent channel flow for reducing skin friction drag. As is often the case with wall turbulence control, we consider wall blowing and suction as a control input, which is defined at each refcomputational grid point on the wall. This makes the degrees of freedom of the control input quite large (O($10^4$)), compared with those assumed in the existing applications of reinforcement learning. This paper organizes as follows: After introducing our problem setting in §2, we explain the framework of the present reinforcement learning in detail in §3. Then, we present new control policies obtained in the present study, and their control results in §4. In §5, we further discuss how the unique features of the present control policies lead to high control performances. Finally, we summarize the present study in §6.

## 2 Problem Setting

### 2.1 Governing equations and boundary conditions

We consider a fully developed turbulent channel flow with wall blowing and suction as a control input as shown in figure 1. The coordinate systems are set so that $x, y$ and $z$ correspond to the streamwise, wall-normal and spanwise directions, respectively. The corresponding velocity components are denoted by $u, v$ and $w$. Time is expressed by $t$. The origin of the coordinate is placed on the bottom wall as shown in figure 1. Unless otherwise stated, we consider only the bottom half of the channel due to the symmetry of the system. The governing equations of the fluid flow are the following incompressible Navier-Stokes and continuity equations:

$$\frac{\partial u_i}{\partial t} + u_j \frac{\partial u_i}{\partial x_j} = -\frac{\partial p}{\partial x_i} + \frac{1}{Re} \frac{\partial^2 u_i}{\partial x_j \partial x_j} \tag{2.1}$$

$$\frac{\partial u_i}{\partial x_i} = 0 \tag{2.2}$$

where $p$ is the static pressure. Throughout this paper, all variables without a superscript are non-dimensionalized by the channel half width $h^*$ and the bulk mean velocity $U_b^*$, while a variable with an asterisk indicates a dimensional value. A constant flow rate condition is imposed, so that the bulk Reynolds number is $Re_b \equiv \frac{2U_b^* h^*}{\nu^*} = 4646.72 *$ where $\nu^*$ is the kinematic viscosity of the fluid. The corresponding friction Reynolds number in the uncontrolled flow is $Re_\tau \equiv \frac{u_\tau^* h^*}{\nu^*} \approx 150$. Here, the friction velocity is defined as $u_\tau^* = \sqrt{\underline{\tau_w^*}/\rho^*}$, where $\rho^*$ is the fluid density, and $\underline{\tau_w^*}$ is the space-time average of the wall friction. Periodic boundary conditions are imposed in the streamwise and spanwise directions. As for the wall-normal direction, we impose no-slip conditions for the tangential velocity components on the wall, while wall blowing and suction with zero-net-mass-flux is applied as a control input:

$$u_i(x, 0, z, t) = \phi(x, z, t)\delta_{i2} \tag{2.3}$$

Here, $\phi(x, z, t)$ indicates the space-time distribution of wall blowing and suction at the bottom wall $(y = 0)$, and $\delta_{ij}$ is the Kronecker delta. Wall blowing and suction is also imposed at the top wall, and its space-time distribution is determined based on the same control policy as that used for the bottom wall, so that the resulting flow is always statistically symmetric with respect to the channel center. The objective of the present study is to find an effective strategy to determine the distributions of $\phi(x, z, t)$for drag reduction.

In reinforcement learning, a control policy (control law) is learned on trial-and-error basis requiring a large number of simulations. In order to reduce the computational cost for the training, we introduce the minimal channel (Jim´enez & Moin 1991), which has the minimum domain size to maintain turbulence. Accordingly, the streamwise, wall- normal and spanwise domain sizes are set to be $(L_x, L_y, L_z) = (2.67, 2.0, 0.8)$. Once a control policy is obtained in the minimal channel, it is assessed in a larger domain with $(L_x, L_y, L_z) = (2.5\pi, 2.0, \pi)$. Hereafter, the latter larger domain is referred to as a full channel

## 2.2 Numerical methodologies

The governing equations (2.1) and (2.2) are discretized in space by a pseudo-spectral method (Boyd 2001). Specifically, Fourier expansions are adopted in the streamwise and spanwise directions, while Chebyshev polynomials are used in the wall-normal direction. For the minimal channel, the number of modes used in each direction is $(N_x, N_y, N_z) = (16, 65, 16)$, whilst they are set to be $(N_x, N_y, N_z) = (64, 65, 64)$for the full channel. The 3/2 rule is applied to eliminate aliasing errors, and therefore the number of grid points in the physical space are 1.5 times the number of modes employed in each direction.

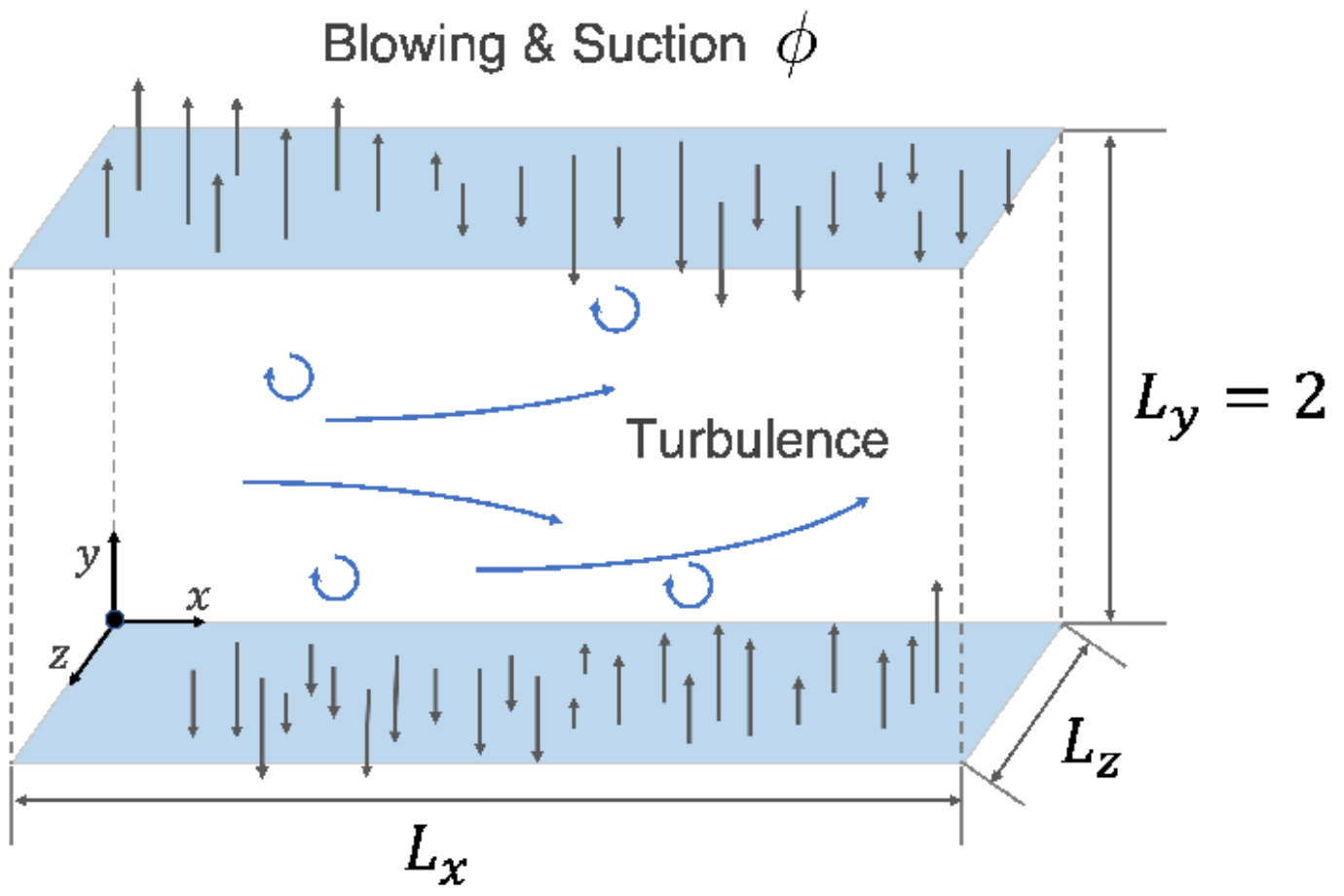

**Figure 1**: Schematic of computational domain and coordinate system

As for the time advancement, a fractional step method (Kim & Moin 1985) is applied to decouple the pressure term from equation (2.1). The second-order Adams-Bashforth method is used for the advection term, while the Euler implicit method for the viscous term. The time step is set to be $\Delta t^+ = 0.06$ and $0.03$ for the minimal and full channels, respectively. The superscript of + denotes a quantity scaled by the viscous scale in the uncontrolled flow throughout this paper. The above setting of the time step ensures that the Courant number is less than unity even with wall blowing and suction. The present numerical scheme has already been validated and successfully applied to control and estimation problems in the previous studies (Yamamoto *et al*. 2013; Suzuki & Hasegawa 2017)

## 3 Reinforcement Learning

### 3.1 Outline

Reinforcement learning is a problem where an agent (learner) learns the optimal policy that maximize a long-term total reward through trials and errors. Specifically, an agent receives a states from an environment (control target) and decides an action a based on a policy $\mu(a|s)$. By executing the action against the environment, the state changes from $s$ to $s'$ and a resulting instantaneous reward r is obtained. Then, $s'$ and r are fed back to the agent and the policy is updated. With the new policy, the next action $a'$ under the new state $s'$ is determined. By repeating the above interaction with the environment, the agent learns the optimal policy. If the next state $s'$ and the instantaneous reward r depend only on the previous state s and the action a, this process is called the Markov Decision Process (MDP), which is the basis of the reinforcement learning (Sutton & Barto 2018).

In the current flow control problem, the environment is a fully developed turbulent channel flow, whereas the state is sensing signal from the instantaneous flow field, and the action corresponds to the control input, i.e., wall blowing and suction. The instantaneous reward $r(t)$ is the friction coefficient $C_f(t)$ with a negative sign, since the reward is defined to be maximized, while the wall friction should be minimized in the present study. Specifically, it is defined as

$$r(t) = -C_f(t) \tag{3.1}$$

where

$$C_f(t) = \frac{\overline{\tau_w}}{\frac{1}{2}\rho U_b^2} \tag{3.2}$$

Here, $\overline{\tau_w}$is the spatial mean of the wall shear stress over the entire wall, and therefore both r and $C_f$ are functions of time as explicitly written in equations (3.1) and (3.2).

Our objective is to find an efficient control policy which describes the relationship be- tween the flow state and the action for maximizing the future total reward. In the present study, we use the Deep Deterministic Policy Gradient (DDPG) algorithm (Lillicrap *et al.* 2015), which is a framework to optimize a deterministic policy. Specifically, this algorithm consists of two neural networks called an actor and a critic as shown in figure 2. The input of the actor is the state s, while its output is the action $a$. Therefore, the actor dictates a control policy $\mu(a|s)$., and it has to be optimized. For this purpose, another network, i.e., a critic, is introduced. The inputs of the critic are the current state s and the action a. The critic outputs the estimation of an action value function $Q^\mu(s,a)$., i.e., the expected total future reward when a certain action a is taken under a certain state s. It should be noted that, during the training, although the instantaneous reward, i.e., instantaneous wall friction, is obtained at every time step from simulation, we generally do not know$Q^\mu(s,a)$, since it is determined by the equilibrium state after the current control policy μ is continuously applied to the flow field. The role of the critic network is to estimate $Q^{\mu}(s,a)$ from past states, actions and resulting rewards.

As for training the networks, the actor is first trained so as to maximize the expected total reward $Q^{\mu}(s,a)$ while fixing the clitic network. Then, the critic is optimized so that the resulting $Q^{\mu}(s,a)$ minimizes the following squared residual of the Bellman equation

$$L_{Critic} = \{r(s,a) + \gamma Q^{\mu}(s',a') - Q^{\mu}(s,a)\}^2 \quad (3.3)$$

As shown in figure 2, the two networks are coupled and trained alternatively, so that both of them will be optimized after a number of trials. Here, $\gamma$ is the time discount rate. If it is set to be small, the agent searches for a control policy yielding a short-term benefit. In contrast, when $\gamma$ approaches to unity, the policy is optimized from a longer- term perspective. Meanwhile, it is also known that, when it is set too large, the agent tends to select no action to avoid failure, i.e., drag increase. In this study, $\gamma$ is set to be 0.99, which is the same as that commonly used in previous studies (Paris *et al*. 2021; Fan *et al*. 2020).

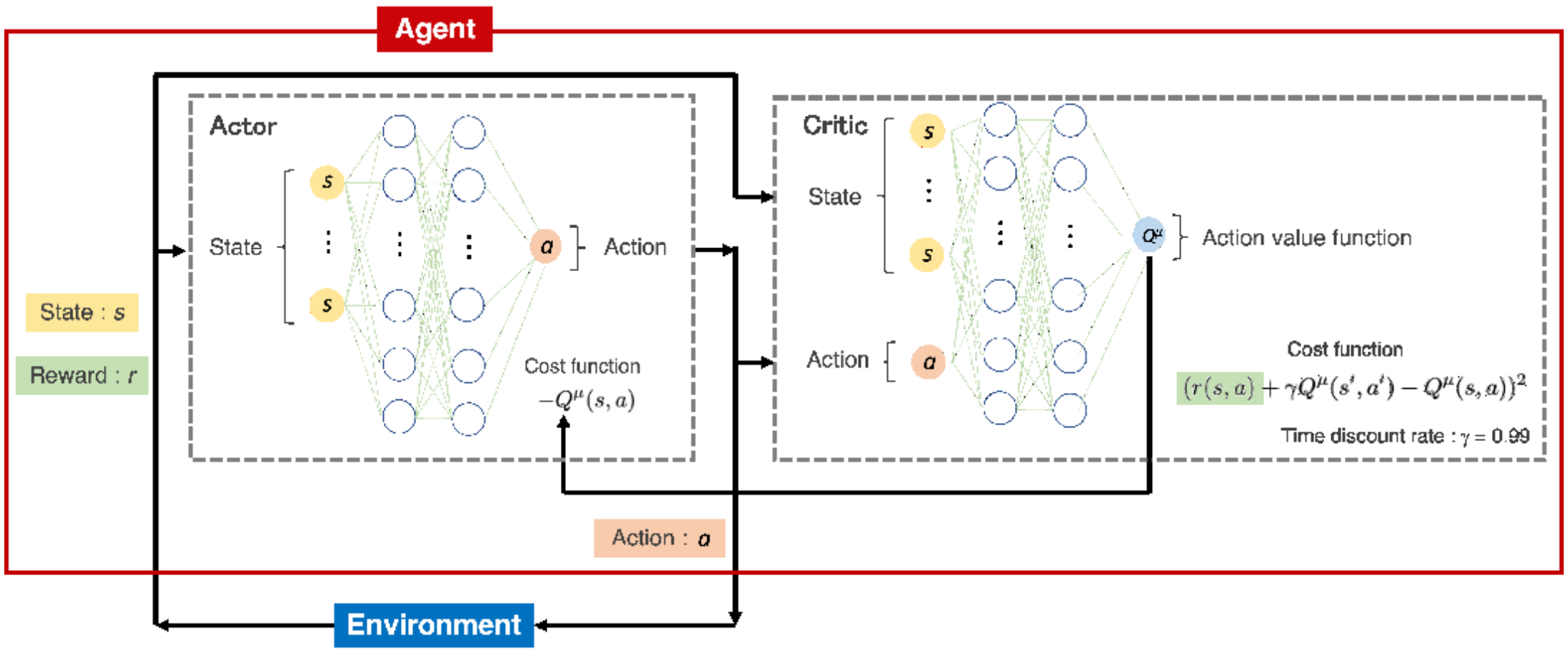

**Figure 2**: Schematic diagram of the DDPG algorithm.

## 3.2 State action, and network setting

Ideally, the velocity field throughout the entire domain should be defined as the state, and wall blowing and suction imposed at each grid point should be considered as the action. In such a case, however, the degree of freedom of the state and the action becomes quite large, so that network training will be difficult. Meanwhile, considering the homogeneity of the current flow configuration in the streamwise and spanwise directions, wall blowing and suction could be decided based on the local information of the flow field. For example, the opposition control (Choi *et al.* 1994), which is one of the well-known control strategies, applies local wall blowing and suction so as to cancel the wall-normal velocity fluctuation above the wall. Hence, in the present study, we also assume that a local control input can be decided solely based on the velocity information above the location where the control is applied. Specifically, we set the detection plane height to $y_d^+ = 15$ ,which is found to be optimal for the opposition control in previous studies (Hammond *et al.* 1998; Chung & Talha 2011). We note that we have conducted additional configuration where the state is defined as the velocity information at multiple locations above the wall. It was found that the resultant control performance is not significantly improved from that obtained in the present configuration with a single sensing location, and the largest weight was confirmed around $y^+ = 15$ (see, Appendix A). Hence, the present study focuses on a control with the single detection plane located at $y_d^+ = 15$ from the wall.

As a first step, we consider the simplest linear actor defined as follows:

$$a \equiv \phi(x, z, t) = \alpha \cdot v'(x, y_d, z, t) + \beta + N \tag{3.4}$$

Here, the prime indicates the deviation from the spatial mean, so that $v' = v - \bar{v}$. α and $\beta$ are constants to be optimized. In order to enhance the robustness of the trainig, a random noise $N$ with zero mean and the standard deviation of 0.1 in a wall unit is added. Throughout the present study, the same magnitude of $N$ is used in all the cases. In the present flow configuration, where periodicity imposed in the streamwise and spanwise directions, $\bar{v}$ is null, and therefore $v' = v$. We also note that the same values of $\alpha$ and $\beta$ are used for all locations on the wall. In addition, a net-mass-flux from each wall is assumed to be zero, so that $\beta$ is zero. Eventually, the above problem reduces to optimizing the single parameter α in the actor. This configuration will be referred to as Case Li00 as shown in table 1. For this control algorithm (3.4), the previous study (Chung & Talha 2011) reported that the optimal value of α is around unity. The purpose of revisiting this configuration is to assess whether the present reinforcement learning can reproduce the opposition control, find the optimal value of α, and achieve a drag reduction similar to that reported in the previous study. We also note that the output of the actor is clipped to $-1 \leq \phi^+ \leq 1$ before applying it to the flow simulation in order to avoid a large magnitude of the control input.

Considering the skin friction drag is directly related to the Reynolds shear stress, $-\overline{u'v'}$ (Fukagata *et al.* 2002), the streamwise velocity fluctuation, $u'$, would also be worth considering in addition to $v'$. Hence, for the rest of the cases shown in table 1, both $u'$ and $v'$ at $y_d^+ = 15$ are considered as the state. The actor network has 1 layer and 8 nodes as shown on the left of figure 3. We have changed the size of the actor network and found that further increases in the numbers of layers and nodes do not improve the resultant control performance (see, Appendix B). The mathematical expression of the present actor network is given as follows:

$$a \equiv \phi(x,z,t) = \tanh[\sigma\{u'(x,y_d,z)\alpha_{11} + v'(x,y_d,z)\alpha_{12} + \beta_1\} \cdot \alpha_2 + \beta_2] + N \qquad (3.5)$$

Here, $\alpha_{11}$, $\alpha_{12}$, $\beta_1$ and $\alpha_2$ are vectors having the same dimension as the number of the nodes, while $\beta_2$ is a scalar quantity. As for the activation function $\sigma$, we consider ReLU, Sigmoid, Leaky ReLU, and hyperbolic tangent, which are respectively referred to as Cases R18, S18, LR18, T18 as listed in table 1. The last two numbers put in each case name represent the numbers of layers and nodes employed in the actor. We also note that a hyperbolic tangent is used for the activation function of the output layer in order to map the range of the control input into $\|\phi^+\| < 1.0$.

The network structure of the critic is schematically shown on the right of figure 3. It consists of 2 layers with 8 and 16 nodes for the first and second layers for the state, and another 1 layer network with 16 nodes for the action. Then, the two networks are integrated by additional two layers with 64 nodes, and the final output is the action value function $Q^\mu(s,a)$. ReLU is used for the activation function.

In order to take into account the cost for applying the control, we extend the reward as follows:

$$r = -C_f - d\frac{\overline{(\phi^+)^2}}{2} \qquad (3.6)$$

The second term of equation (3.6) represents the cost of control and d is a weight coefficient that determines the balance between the wall friction and the cost for applying the control. In the present study, d is systematically changed from 0 to 0.1, which are referred to as R18, R18D1-D3 (see, table 1).

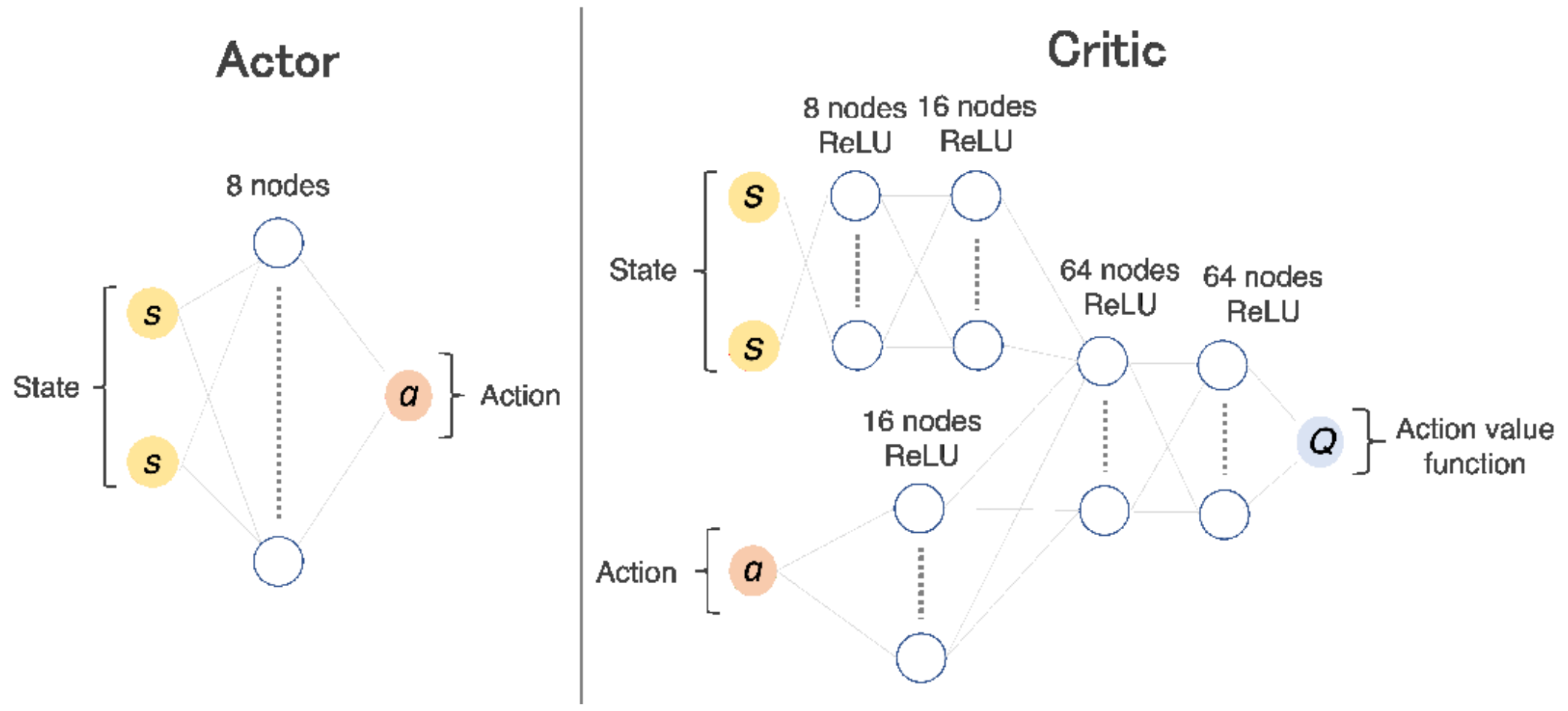

**Figure 3**: Network structures of the actor (left) and the critic (right)

| Case | State | Layers | Nodes | Activation function | $d$ |
|---|---|---|---|---|---|
| Li00 | $v'\|_{y^+=15}$ | 0 | 0 | None | 0 |
| R18 | $u', v'\|_{y^+=15}$ | 1 | 8 | ReLU | 0 |
| S18 | $u', v'\|_{y^+=15}$ | 1 | 8 | Sigmoid | 0 |
| LR18 | $u', v'\|_{y^+=15}$ | 1 | 8 | LeakyReLU | 0 |
| T18 | $u', v'\|_{y^+=15}$ | 1 | 8 | tanh | 0 |
| R18D1 | $u', v'\|_{y^+=15}$ | 1 | 8 | ReLU | 0.01 |
| R18D2 | $u', v'\|_{y^+=15}$ | 1 | 8 | ReLU | 0.05 |
| R18D3 | $u', v'\|_{y^+=15}$ | 1 | 8 | ReLU | 0.1 |

**Table 1**: Considered cases with the corresponding state, numbers of layers and nodes, an activation function, and the weight coefficient $d$ for the control cost.

### 3.3 Learning procedures

Figure 4 shows the general outline of the present learning procedures. The two net- works, i.e., actor and critic, are trained in parallel with flow simulation within a fixed time interval, which is called an episode. In the present study, the episode duration is set to be $\mathrm{T}^+ = 600$, and the flow simulation is repeated within the same interval, i.e., $t \in [0, T]$ In each episode, the flow simulation is started from the identical initial field at $t = 0$, which is a fully developed uncontrolled flow. For $t > 0$, the control input $\phi$ is applied from the two walls in accordance with the control policy $\mu(a|s)$. We set the episode duration as$\mathrm{T}^+ = 600$, so that the period covers the entire processes in which the initial uncontrolled flow transits to another fully developed flow with the applied control. If the episode length is too short, the flow does not converge to a fully developed state, so that the obtained policy is only effective for the initial transient after the onset of  the control. Meanwhile, if the episode length becomes longer, the obtained policy is more biased to the fully developed state under the control, and therefore might not be effective for the initial transient. According to our experiences, the episode duration should be determined so that it covers the entire procedures for the initial uncontrolled flow to converges to another fully developed state after the onset of a control. Of course, the transient period should generally depend on a control policy and also a flow condition, and therefore the optimal episode duration has to be found by trials and errors.

Within each episode, the agent consecutively interacts with the flow by applying a control, and receives the instantaneous reward (3.1 or 3.6). Based on each interaction, the networks of the actor and the critic are updated. The Adam optimizer is used for both the networks, whereas the learning rate is set to be 0.001 and 0.002 for the actor and the critic, respectively. The buffer size is set to be 5,000,000, while the batch size is 64. In the present study, the networks are trained every short interval of $\Delta t^+_{update} = 0.6$. Accordingly, the control input is also recalculated from the updated policy at the same time interval. Within the time interval, the control input is linearly interpolated (see, figure 4). Ideally, a smaller time interval is better, since there will be more chances to update the networks. Meanwhile, it is known that a short training interval often causes numerical instability (Rabault *et al.* 2019; Fan *et al.* 2020). Our preliminary simulation results indicate that $\Delta t^+_{update} = 0.6$ leads to the best control performance.

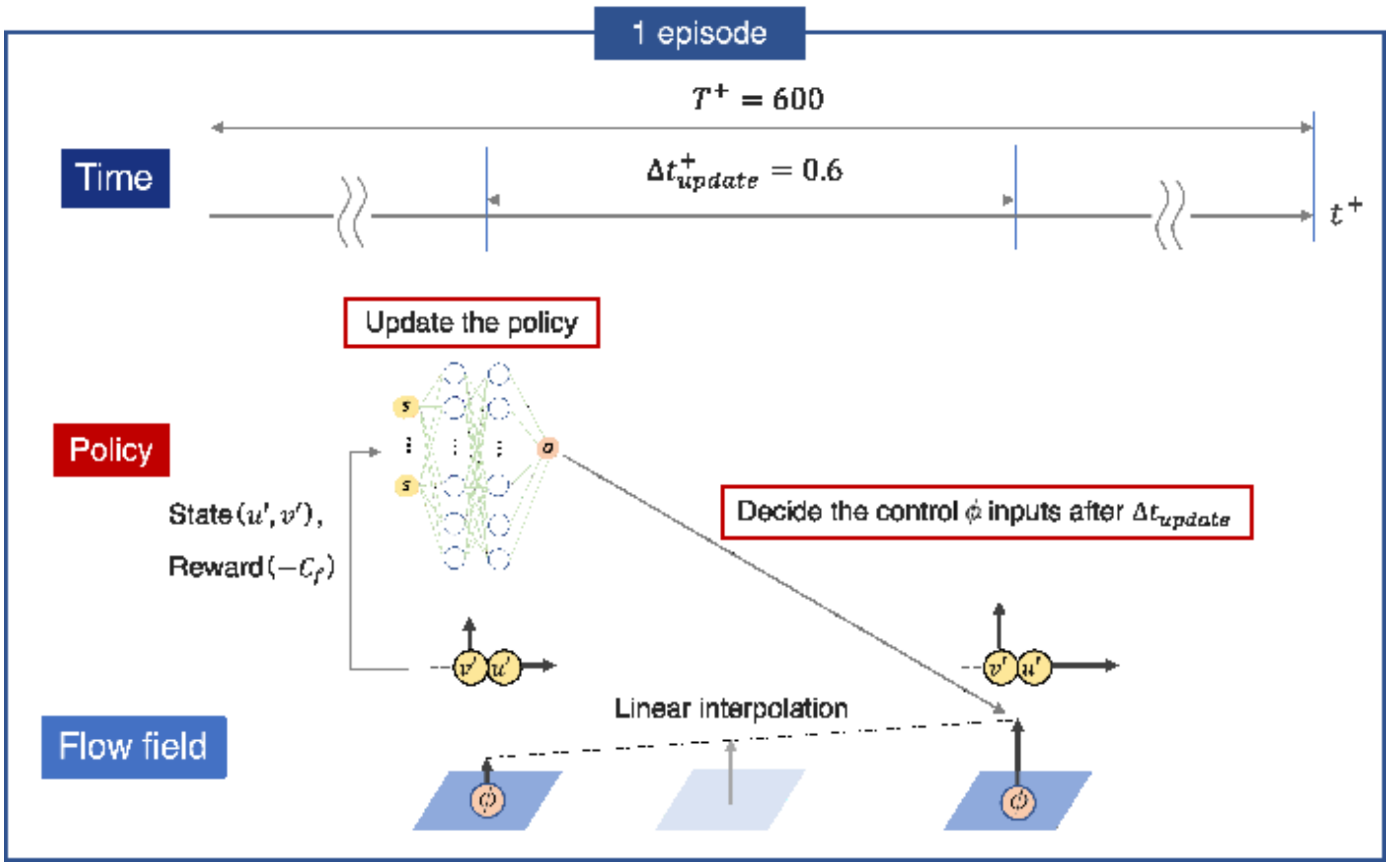

**Figure 4**: Schematic of learning process

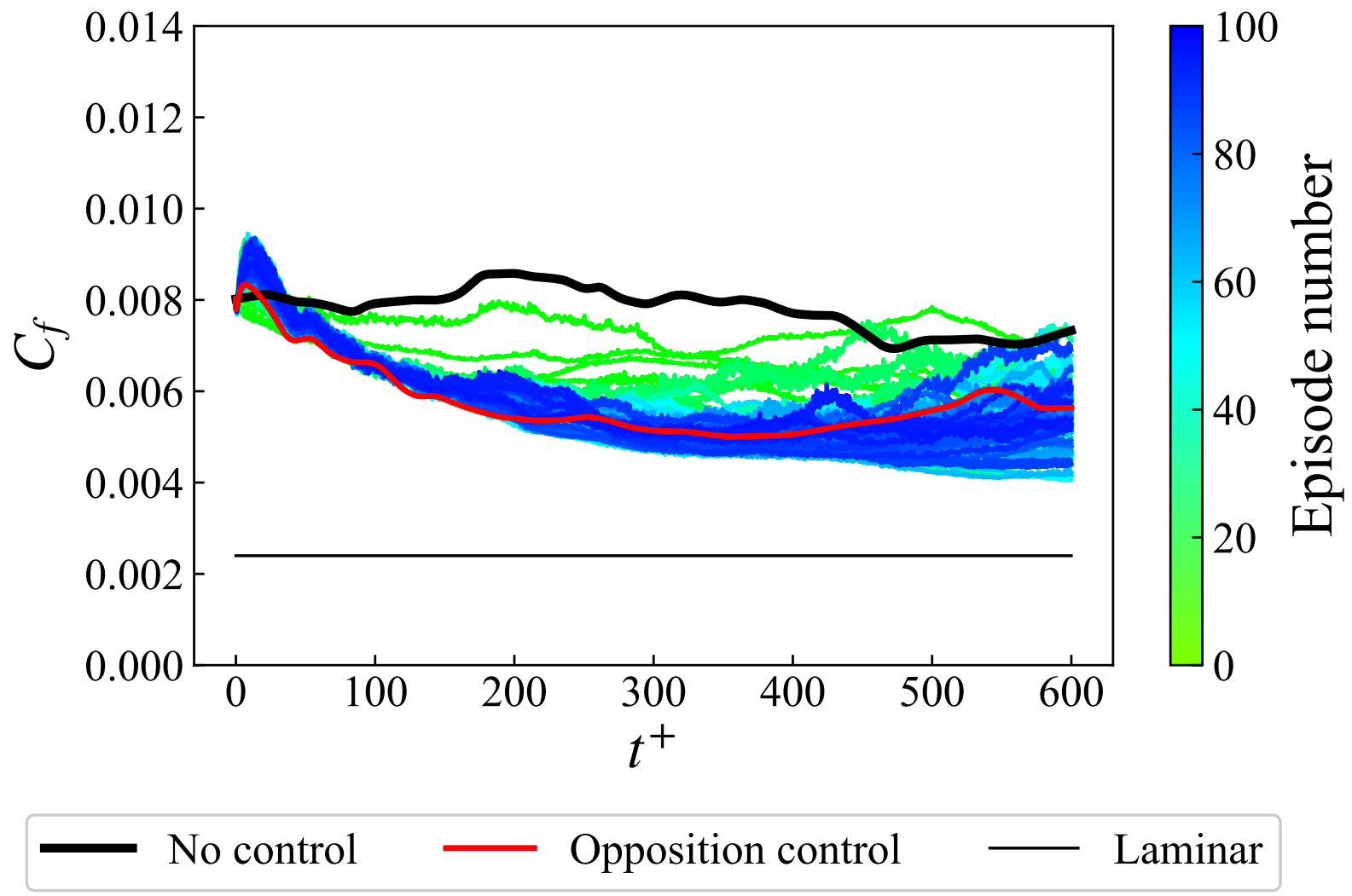

**Figure 5**: Temporal evolution of $C_f$ obtained in each episode for Case Li00. With increasing the episode number, the line color changes from green to blue. Black and red lines correspond to those for uncontrolled and opposition control cases

## 4 Results of Reinforcement Learning

### 4.1 Linear policy: revisit to the opposition control

As a first step, we consider Case Li00, where only the wall-normal velocity fluctuation $v'$ at the detection plane at $y_d^+ = 15$is used as a state, and the policy dictating the rela- tionship between the state and the control input is linear as described in equation (3.4). The time traces of the instantaneous $C_f$ for different episodes are shown in figure 5. The line color changes from green to blue as the number of episodes increases. For comparison, we also plot the temporal evolution of $C_f$ for the uncontrolled and opposition control cases with black and red lines, respectively. It can be seen that $C_f$ is successfully reduced as the training proceeds, and eventually converges to a value similar to that obtained by the opposition control.

In figure 6, the time average $\langle C_f \rangle$ of the instantaneous friction coefficient is shown, where the bracket $\langle \cdot \rangle$ indicates the time average within the final period of $500 \leq t^+ \leq 600$ in each episode. It can be seen that $\langle C_f \rangle$ decreases for the first ten episodes, and then converges to the value obtained by the opposition control. Figure 7 shows the policy, i.e., the control input versus the state, obtained at the end of each episode. The line color changes from green to blue with increasing the episode. As described by equation (3.4), the relationship between the state $v'$ and the control input $\phi$ is linear, and the maximum absolute value of $\phi^+$ is clipped to unity. The red line corresponds to the case of $(\alpha, \beta) = (-1.0, 0)$ in equation (3.4), which was found to be optimal for the opposition control in Chung & Talha (2011). It can be seen that the present policy reproduces the opposition control with the optimal values of $(\alpha, \beta) = (-1.0, 0)$ quite well, while the present policy has a slightly steeper slope. This is probably attributed to the fact that the magnitude of $\phi$ is clipped in the present policy. From the above results, we validate that the present reinforcement learning successfully finds the optimal linear control policy which has been reported in the previous studies (Choi *et al.* 1994; Chung & Talha 2011)

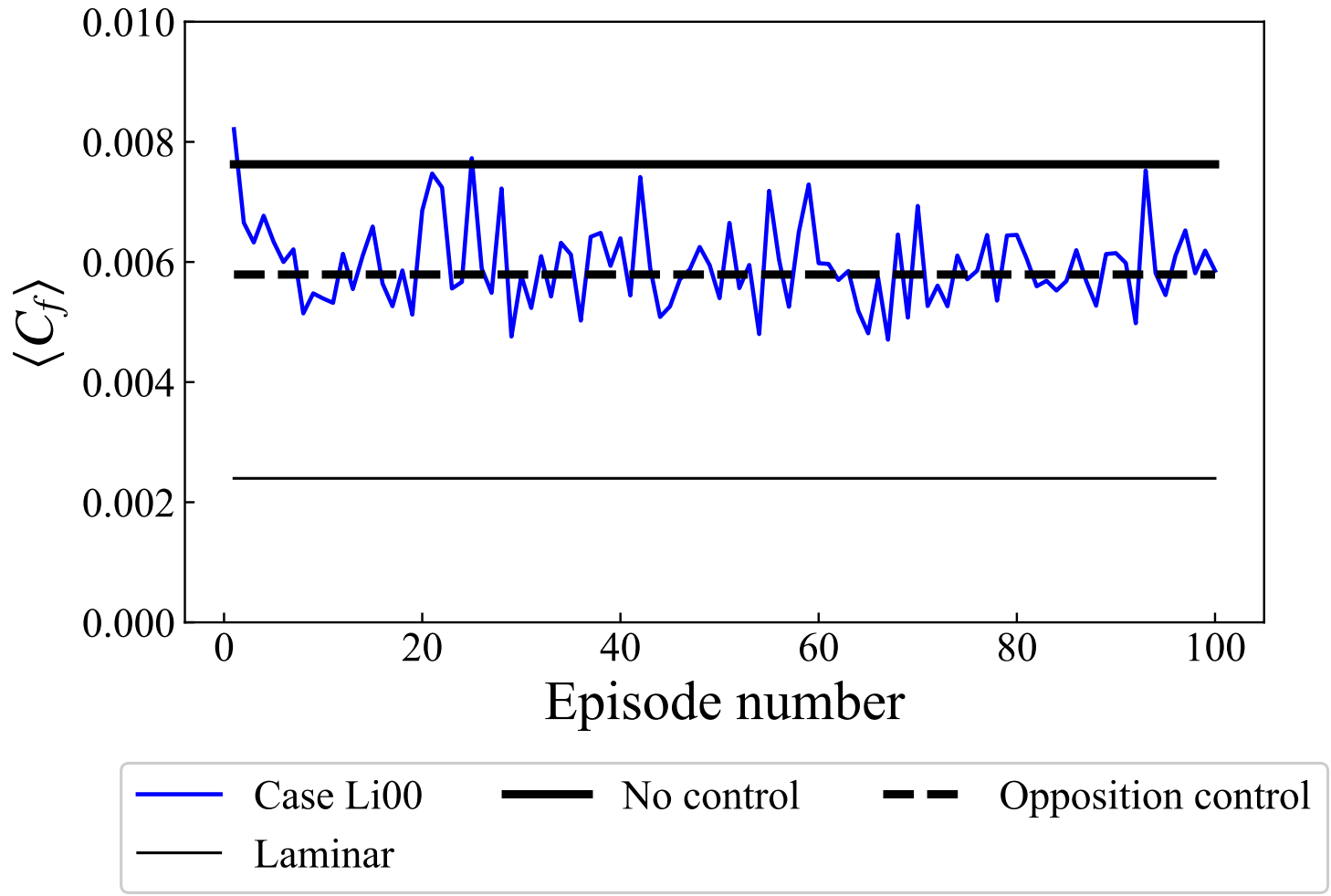

**Figure 6**: Time average of the friction coefficient $\langle C_f \rangle$ at the final period of $500 \leq t^+ \leq 600$ in each episode. Blue: Case Li00, thick black: uncontrolled, black broken: opposition control, thin black: laminar

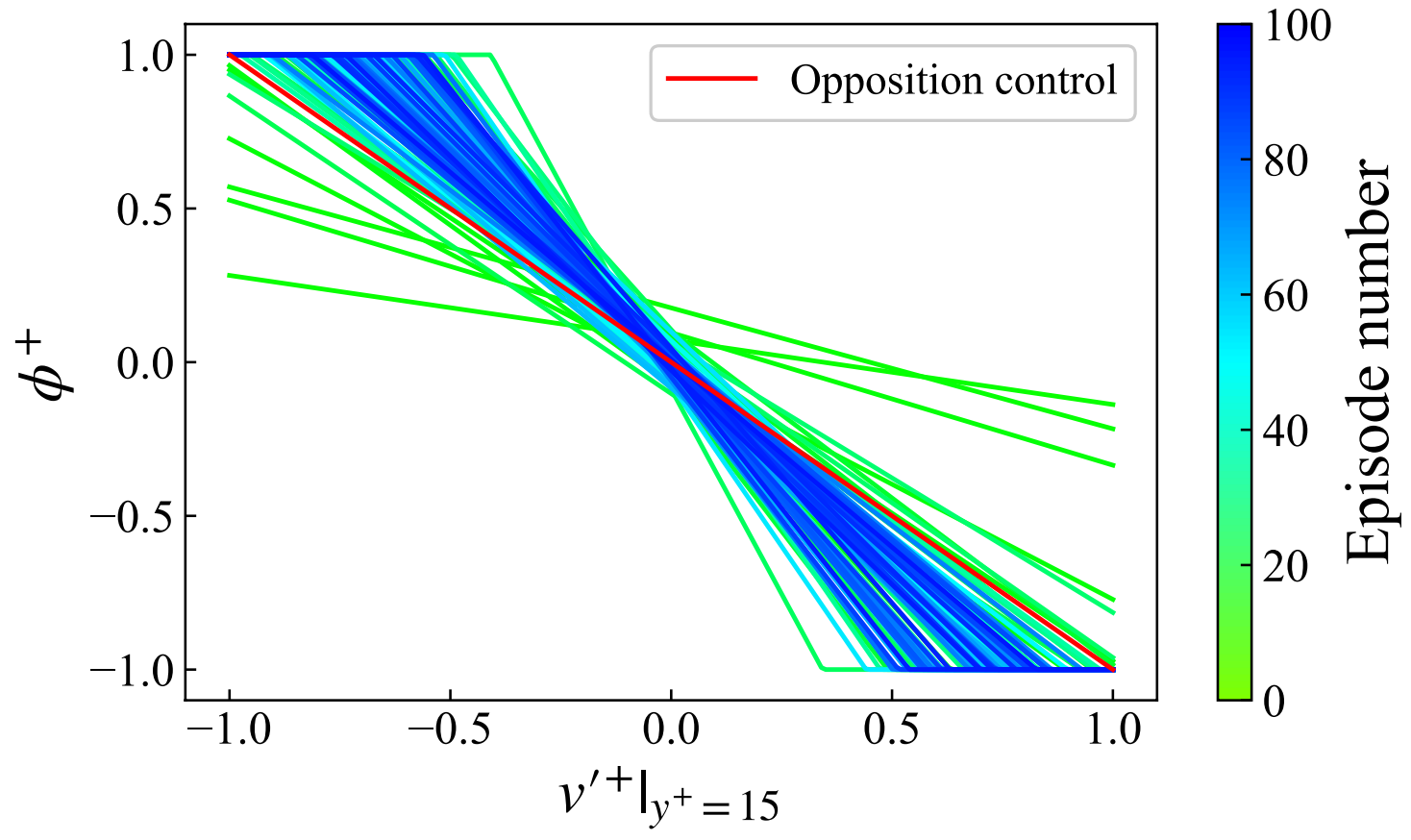

**Figure 7**: Obtained policy in Case Li00 at the end of each episode. With increasing the episode number, the line color changes from green to blue. The red line represents the opposition control where $(w, \beta) = (-1.0,0)$

## 4.2 Nonlinear control policies

In this subsection, we present the results obtained by non-linear policies, where a hidden layer and a nonlinear activation function are added to the actor network as listed in table 1.

### 4.2.1 Obtained policies

Figure 8 shows $\langle C_f \rangle$ as a function of the episode number for Cases R18, S18, LR18, and T18 using different activation functions. For all the cases, $\langle C_f \rangle$ reduces from the uncontrolled value with increasing the episode number, and eventually converges to a value similar to or even smaller than that achieved by the opposition control. Especially, higher drag reduction rates than that of the opposition control can clearly be confirmed in Cases R18, LR18 and S18.

The policy obtained at the best episode where the maximum drag reduction rate is achieved in each case is shown in figure 9 (b)-(e), where the control input φ is plotted as a function of the state $(u', v')$ at $y_d^+ = 15$. Red and blue colors correspond to wall blowing and suction, respectively. For reference, we also plot the policy of the opposition control defined by equation (3.4) with $(\alpha, \beta) = (-1.0, 0)$ in figure 9 (a). In this case, the control input depends on only $v'$, so that the color contours are horizontal, and the control input $\phi$ linearly depends on the state $v'$.

In contrast, the present non-linear control policies shown in figure 9 (b)-(e) obviously depend on not only $v'$, but also $u'$. In addition, the control input rapidly switches between wall blowing and suction drastically depending on the state, i.e., $u'$ and $v'$ at $y_d^+ = 15$. Specifically, for Cases R18 and T18 shown in figure 9 (b) and (e), respectively, the boundary between wall blowing and suction is inclined, so that wall blowing is applied when a high-speed fluid ($u' > 0$) approaches to the wall ($v' < 0$)), while wall suction is applied for upwelling ($v' < 0$)) of low-momentum fluid ($u' < 0$). On the other hand, for Cases S18 and LR18 shown in figure 9 (c) and (d), respectively, the boundary between wall blowing and suction is almost vertical, so that the control input mostly depends on the streamwise velocity fluctuation $u'$ only. It should be emphasized that such complex non-linear relationships between the state and the control input can be first obtained by introducing the neural network for the actor. We also plot the joint probability density function (PDF) of $u'$ and $v'$ at $y_d^+ = 15$for the uncontrolled flow in figure 9 (f). It can be confirmed that the joint PDF roughly fits in the plot range, and the boundaries between wall blowing and suction obtained in all the cases cross the central part of the joint PDF

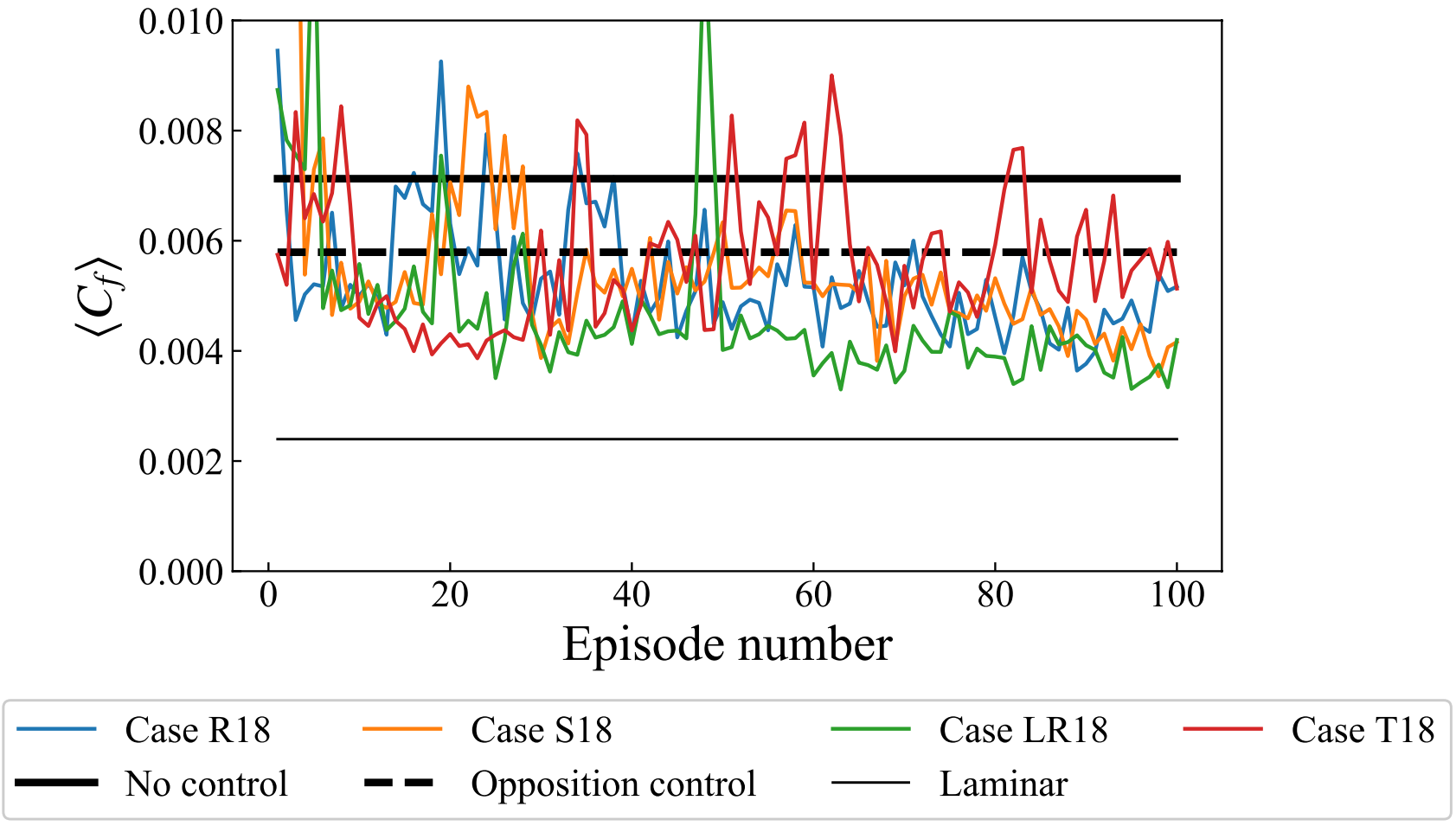

**Figure 8**: $\langle C_f \rangle$ versus the episode number for different policies obtained in the present reinforcement learning. blue: Case R18, yellow: Case S18, green: Case LR18, red: Case T18.

### 4.2.2 Control performance of obtained policies

As mentioned in section 3.3, the present control policies are obtained through iterative training within the fixed episode period of $T^+ = 600$. In order to evaluate their control performances, we continue applying the obtained control policies for a longer time period in the minimal channel. The time evolutions of the instantaneous $C_f$ for the obtained policies are shown in figure 10. It can be seen that all the non-linear policies obtained in the present study achieve higher drag reduction rates than that achieved by the opposition control. Especially, relaminarization can be confirmed in Cases R18 and T18. It should be noted that, however, these policies may not be always optimal, since the control performance of each policy could depend on an initial condition, especially for the minimal channel considered here. Indeed, when we apply the present policies to another initial condition, the resultant drag reduction rates are commonly larger than that obtained by the opposition control, while the relaminarization is not always confirmed (not shown here). Due to the the small domain size of the minimal channel, the turbulent flow becomes intermittent even in the uncontrolled flow (Jim ́enez & Moin 1991), and therefore it is difficult to distinguish whether relaminarization is caused by the applied control or the intermittency of the flow.

In order to evaluate the control performances of the obtained policies, we apply them to the full channel. The results are shown in figure 11. Although relaminarization is no longer achieved in the full channel, it can be seen that the present control policies still outperform the opposition control. Specifically, the drag reduction rates achieved by Cases R18, S18, LR18 and T18 are respectively 31%, 35%, 35% and 27%, while that of the opposition control remains 23%. In summary, it is demonstrated that the control policies obtained in the minimal channel still work in the full channel, and the present reinforcement learning successfully finds more efficient control policies than the existing opposition control.

Before closing this subsection, we also briefly address the generality of the present results. The non-linear policies shown in figure 9 (b-e) commonly exhibit a rapid switch between wall blowing and suction, which may cause numerical oscillations and affect the resultant control performances, especially when a pseudo-spectral method is used. Therefore, we have also assessed the obtained policies in the same flow configurations with another code based on a finite difference method. We found that the resultant drag reduction rates are hardly affected by changing a numerical scheme. The detailed comparisons between the two numerical schemes are summarized in Appendix C.

### 4.2.3 Effects of the control cost

Here, we assess the impacts of the weight d for the control cost in the reward (3.6) by comparing Cases R18, R18D1, R18D2, and R18D3. Figure 12 shows the average drag reduction rates during the final 20 episodes after the flow reaches an equilibrium state for each case in the minimal channel. Specifically, 38%, 38%, 10 % and 7% of drag reduction are obtained in Cases R18, R18D1, R18D2, and R18D3, respectively. We also note that these values change to 27%, 23%, 7% and 14% in the full channel, respectively. From the above results, it can be confirmed that the resulting drag reduction rate decreases with increasing the weight $d$ for the control cost. This suggests that the control cost is properly reflected in the learning process of the present reinforcement learning.

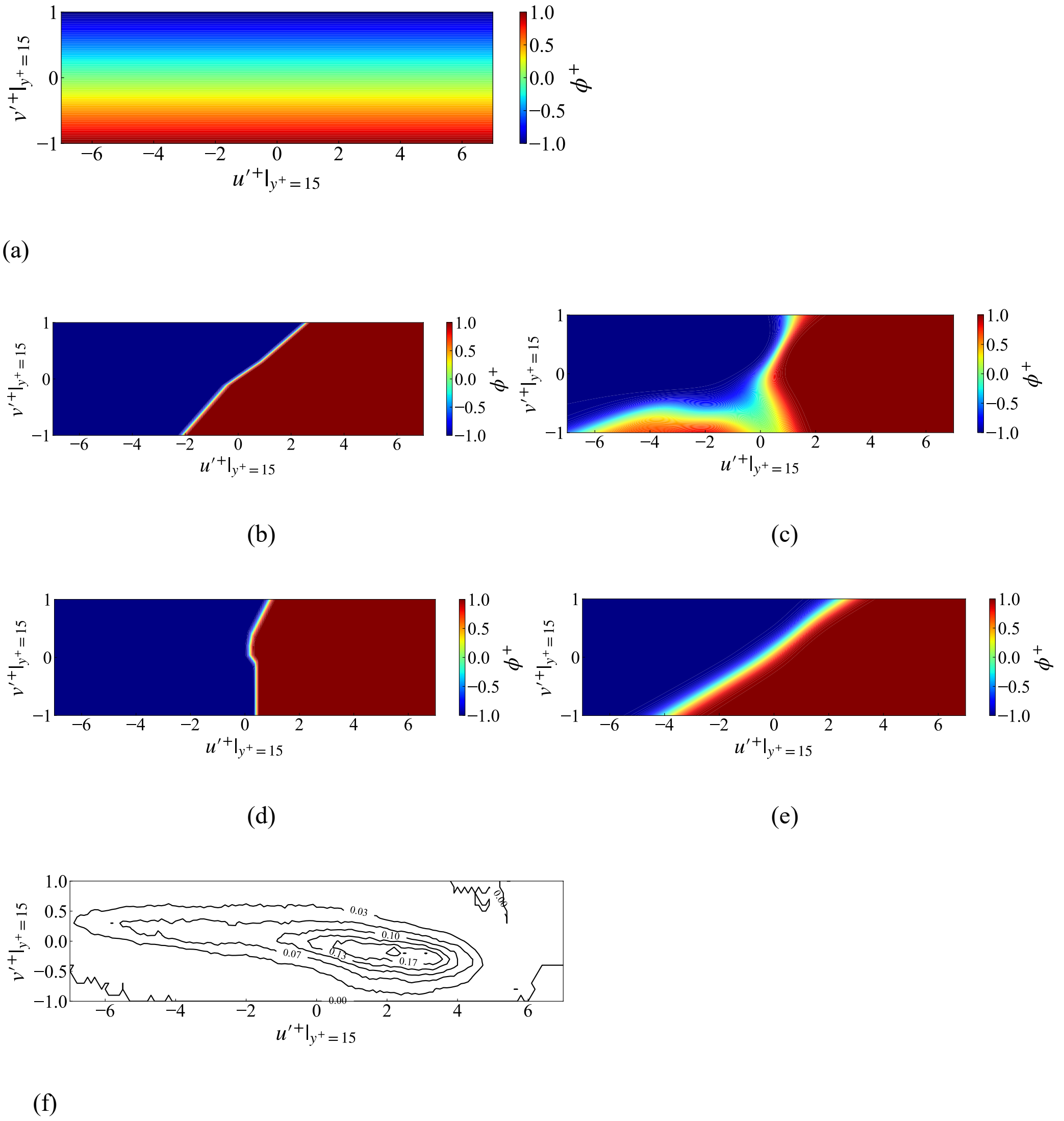

**Figure 9**: Control input as a function of the flow state at $y_d^+ = 15$ (a): opposition control, (b) Case R18, (c) Case S18, (d) Case LR18, (e) Case T18, (f): Joint PDF of $u'$ and $v'$ at $y_d^+ = 15$ in the uncontrolled flow.

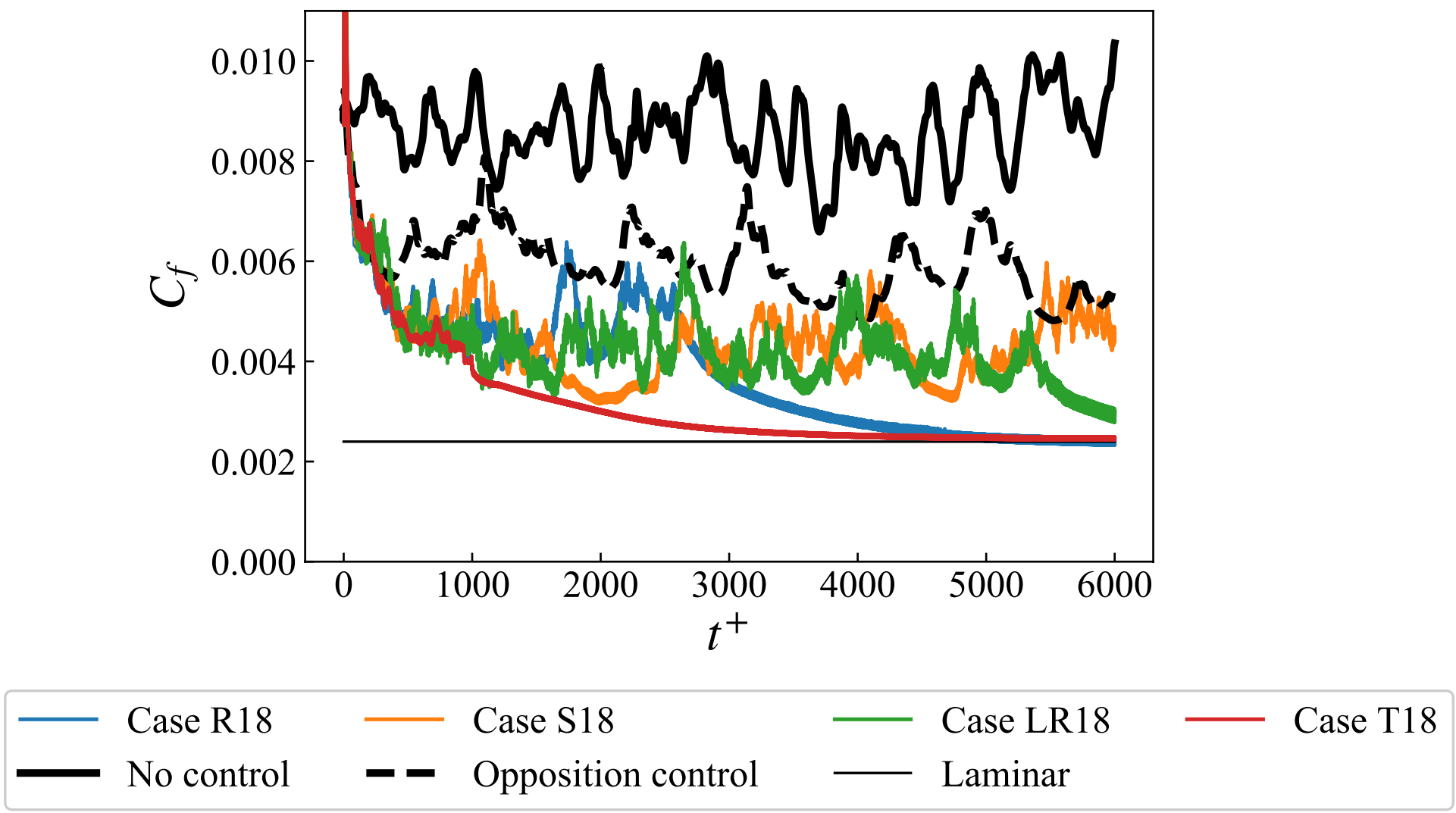

**Figure 10**: Time evolutions of $C_f$ obtained with different policies in the minimal channel

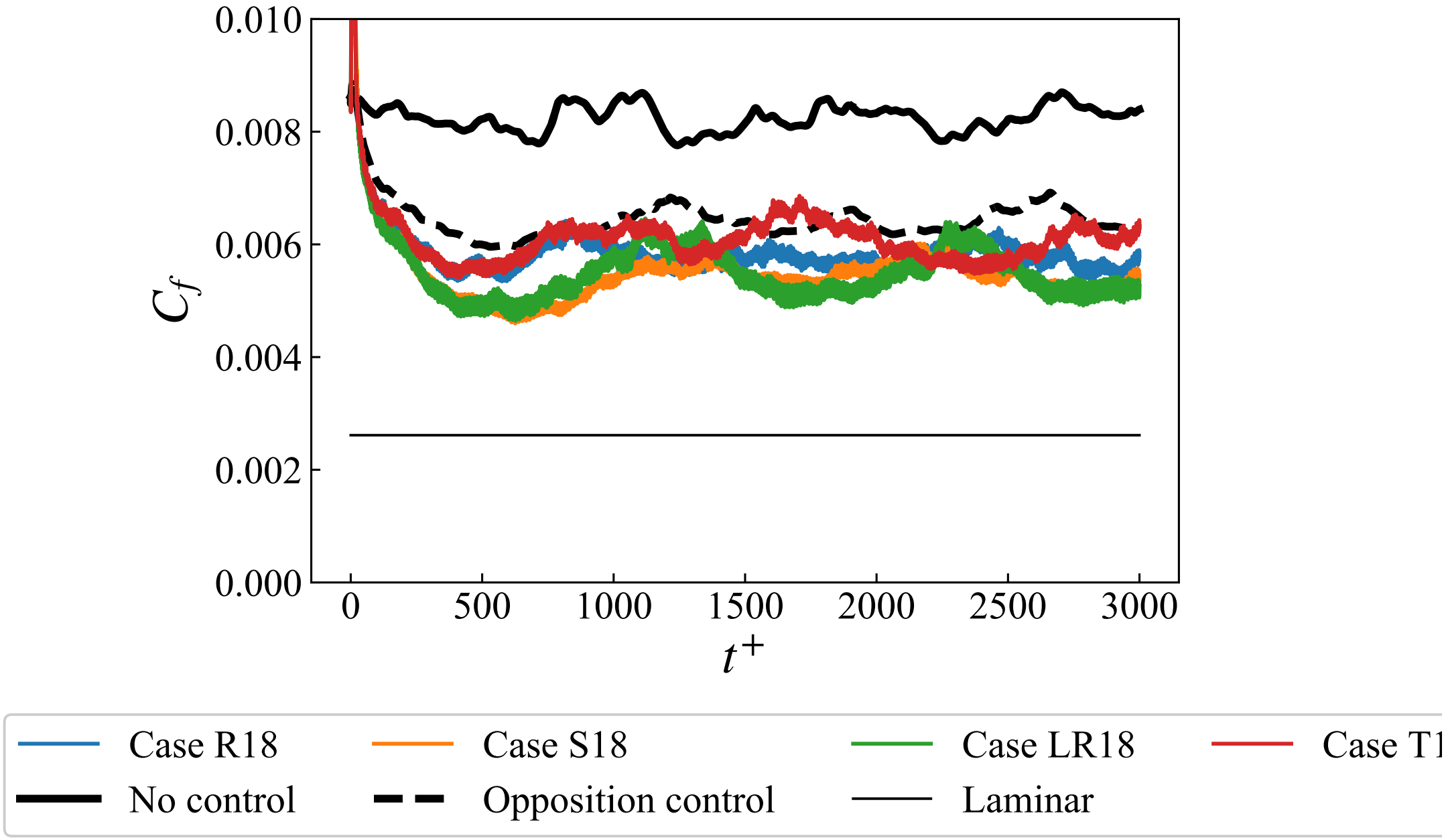

**Figure 11**: Time evoluations of $C_f$ obtained with different policies in the full channel.

The obtained policy at the final episode in each case is shown in figure 13 (b-e) together with that of the opposition control in figure 13 (a). Specifically, in Case R18D1, where d is relatively small, the obtained policy shown in figure 13 (c) is similar to that in Case R18 shown in figure 13 (b), where no control cost is taken into account. It should also be noted, however, that the control input in Case R18D1 almost vanishes in the central region of figure 13 (c). This indicates that, when the cost for the control is relatively small, the obtained policy avoid applying the control when the streamwise and wall- normal velocity fluctuations are relatively small. This is reasonable, since larger velocity fluctuations should have larger contributions to the momentum transfer in the near-wall region. When the cost for the control becomes larger in Cases R18D2 and R18D3, it can be seen that the obtained control policies shown in figures 13 (d) and (e) tends to be similar to the opposition control shown in figure 13 (a). From these results, the opposition control can be considered optimal when the weight for the cost of the control becomes large.

# 5 Feature Analyses of Obtained Policies and Control Inputs

## 5.1 Effects of the rate of the change from wall blowing to suction

The unique features of the control policies obtained in the present reinforcement learning are the rapid switch between wall blowing and suction, and their dependency on the streamwise and wall-normal velocity fluctuations at the detection plane of $y_d^+ = 15$ as shown in figure 9. In this section, we clarify how each feature affects the resulting drag reduction rate, and leads to a higher drag reduction than that obtained by the opposition control. For this purpose, we extract their features, systematically change parameters characterizing them, and evaluate the resulting control performances. We note that all results presented in this section are obtained in the full channel.

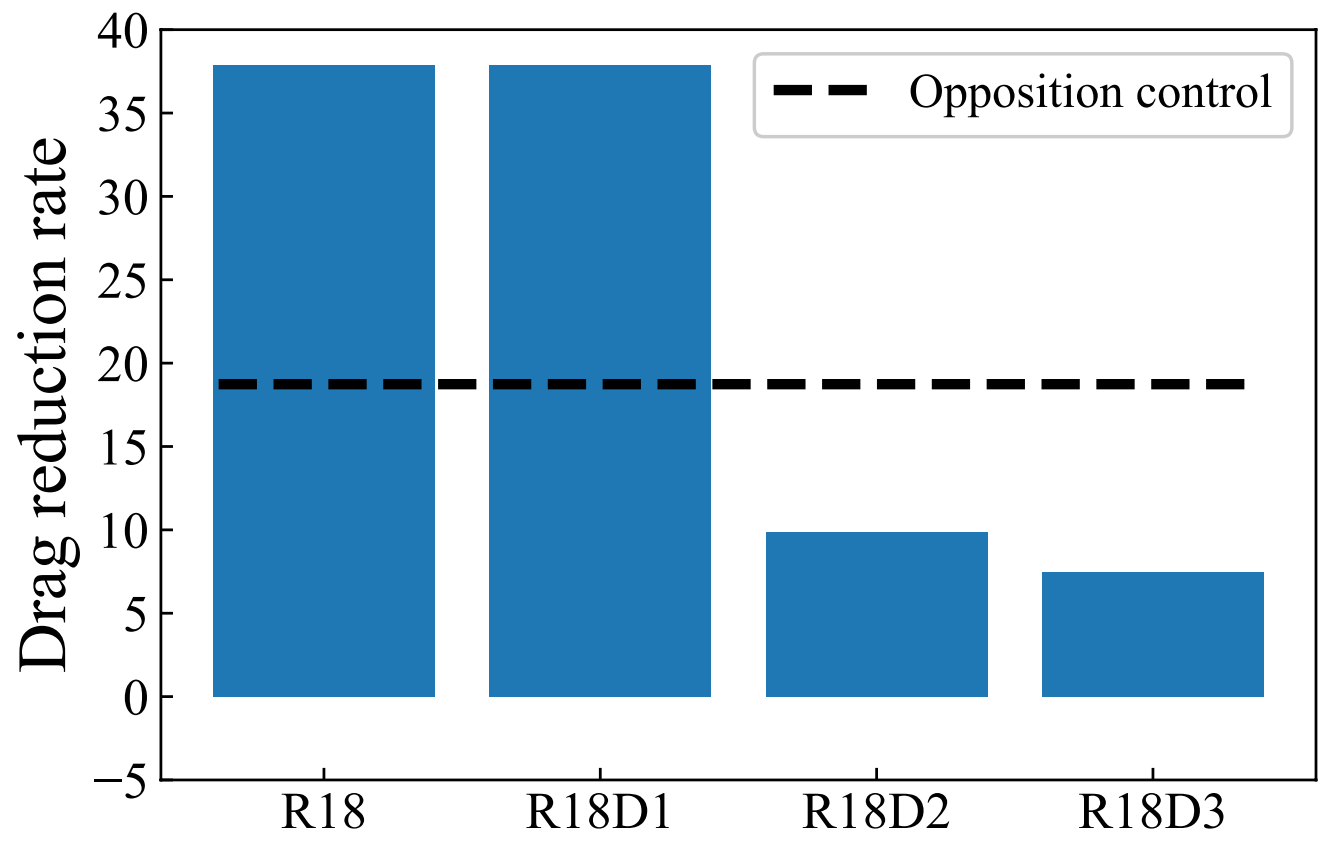

**Figure 12:** Drag reduction rates averaged over the final 20 episodes after the flow reaches an equilibrium state for the minimal channel in Cases R18, R18D1, R18D2 and R18D3. The dashed line corresponds to the drag reduction rate of the opposition control

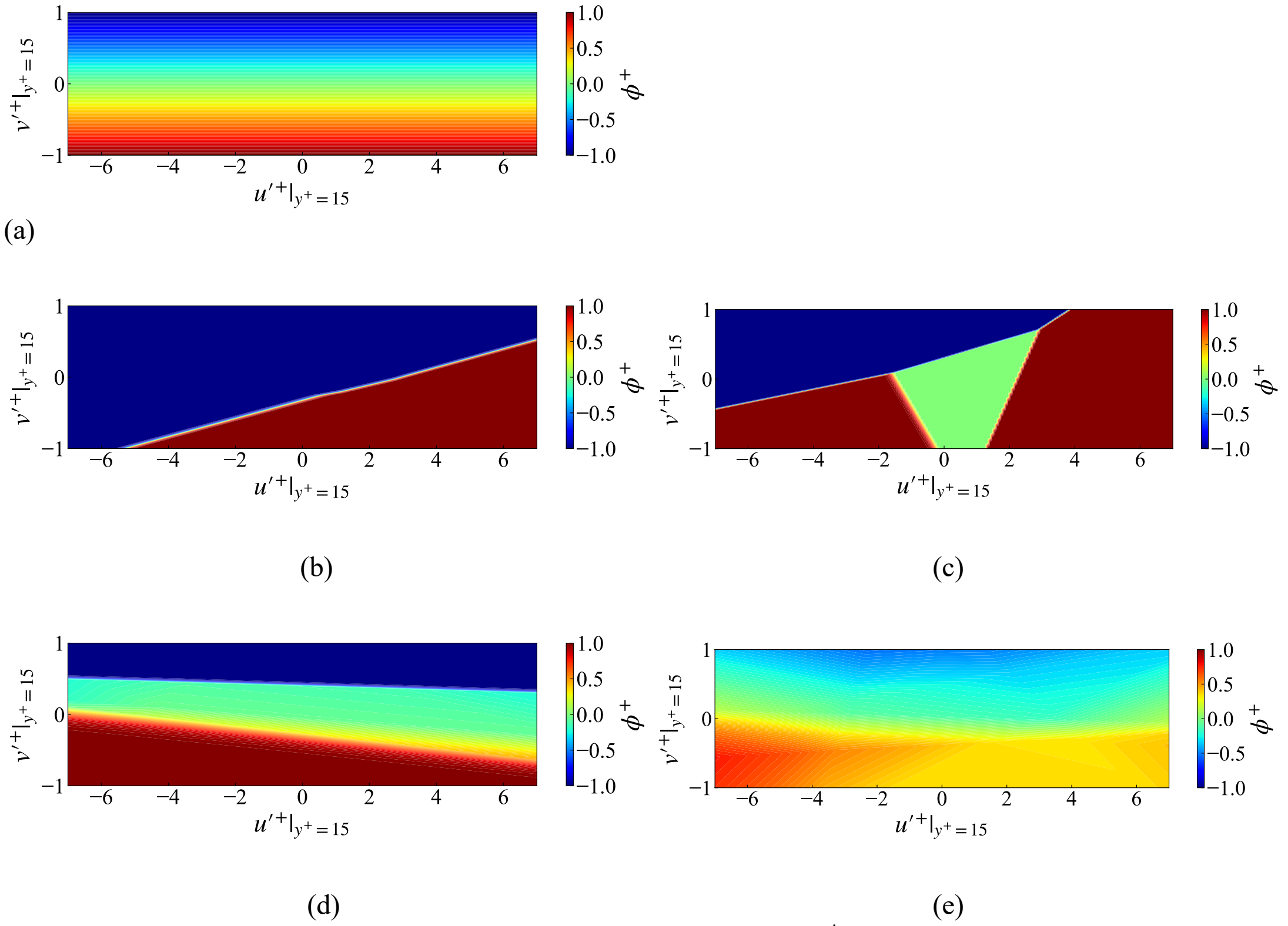

**Figure 13:** Control input as a function of the flow state at $y_d^+ = 15$: (a) Opposition control, (b) Case R18, (c) Case R18D1, (d) Case R18D2, (e) Case R18D3

### 5.1.1 $u'$-based control

We first consider the policies obtained in Cases S18 and LR18 shown in figures 9 (c) and (d). Both of these policies mainly depend on $u'$. In order to clarify how the rate of change from wall blowing to suction in $u'$-based control affects the control performance

| Case | Control policy | $\alpha_u$ | Drag reduction rate |
|---|---|---|---|
| U1 | 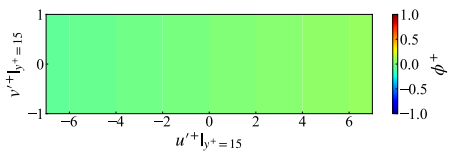 | 0.01 | 6% |
| U2 | 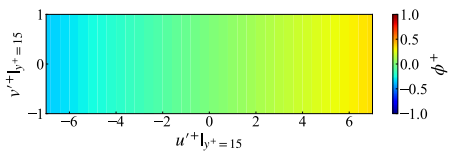 | 0.05 | 16% |
| U3 | 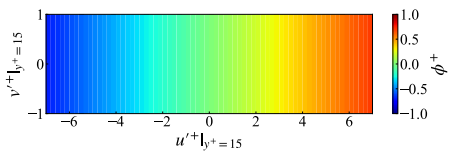 | 0.1 | 20% |
| U4 | 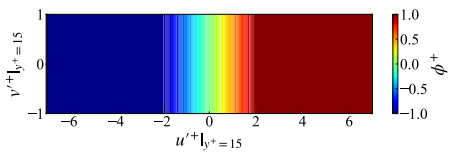 | 0.5 | 30% |
| U5 | 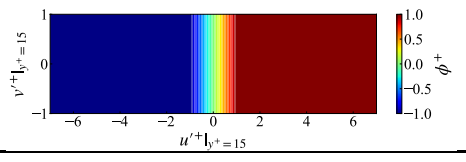 | 1 | 32% |
| U6 | 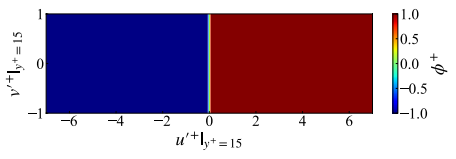 | ∞ | 37% |

**Table 2:** Control policies and resulting drag reduction rates in the full channel for different slopes $\alpha_u$ from wall blowing to suction in $u'$-based control

we consider the following policies:

$$\phi^+ = \begin{cases} \alpha_u u'^+|_{y_d^+=15} & (-1 \leq \alpha_u u'^+ \leq 1) \\ -1 & (\alpha_u u'^+ < -1) \\ 1 & (\alpha_u u'^+ > 1) \end{cases} \tag{5.1}$$

where $\alpha_u$ is a parameter controlling the rate of change from wall blowing to suction, and it is systematically changed from 0.01 to ∞ in the present study. As in the previous cases, $\phi^+$ is constrained from −1.0 to 1.0. The corresponding policies in $u' - v'$ plane and the obtained drag reduction rates are summarized in table 2.

In Case U3, where the slope from wall blowing to suction is moderate, i.e., $\alpha_u = 0.1$, 20% drag reduction rate is achieved, and this value is similar to that obtained by the opposition control. When the rate of the change from wall blowing to suction becomes steeper, i.e., $\alpha_u > 0.1$, the drag reduction rate further increases. From this result, it can be concluded that the sharp change from wall blowing to suction is effective in the $u'$-based control.

### 5.1.2 $v'$-based control

Here, we consider the effects of the rate of change from wall blowing to suction in $v'$-based control. In this case, the policy depends on the wall-normal velocity fluctuation

| Case | Control policy | $\alpha_v$ | Drag reduction rate |
|---|---|---|---|
| Opposition control | 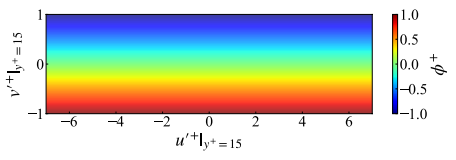 | -1 | 6% |
| V1 | 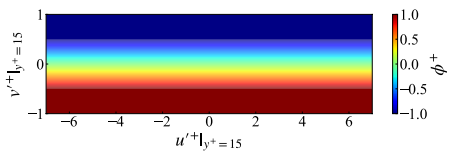 | -2 | 16% |
| V2 | 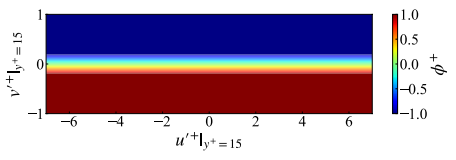 | -5 | 20% |
| V3 | 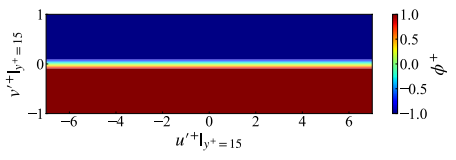 | -10 | 30% |
| V4 | 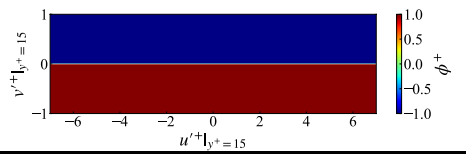 | ∞ | 32% |

| Table 3: | Control policies and resulting drag reduction rates in the full channel for different slopes $\alpha_v$ from wall blowing to suction in $v'$-based control |
|---|---|

$v'$only and it can be expressed as

$$\phi^+ = \begin{cases} \alpha_u v'^+|_{y_d^+=15} & (-1 \le \alpha_u v'^+ \le 1) \\ -1 & (\alpha_u v'^+ < -1) \\ 1 & (\alpha_u v'^+ > 1) \end{cases} \tag{5.2}$$

Again, $\alpha_V$ determines the slope from wall blowing to suction. It should be noted that, when $\alpha_v = -1.0$ it corresponds to the opposition control. The results are summarized in table 3.

In contrast to $u'$-based control summarized in table 2, the opposite trend can be seen. Namely, the drag reduction rate is reduced as $\alpha_V$ increases. It should be noted that Chung & Talha (2011) conducted a parametric survey changing $\alpha_V$ from −0.1 to −1.0, and reported that $\alpha_v = -1.0$is optimal within the range. Hence, we do not repeat these cases here. The current results indicate that the further decrease of $\alpha_v$ from −1.0 does not improve the control performance.

5.1.3 $u'v'$-based control

Next, we consider the effects of the rate of change from wall blowing to suction for a policy which depends on both $u'$ and $v'$. In this case, the considered policies are expressed as follows:

$$\phi^+ = \begin{cases} \alpha_{uv} v'^+|_{y_d^+=15} \left( \dfrac{u'^+|_{y_d^+=15}}{2} - v'^+|_{y_d^+=15} \right) & \left( -1 \le \alpha_{uv} \left( \dfrac{u'^+}{2} - v'^+ \right) \le 1 \right) \\ -1 & \left( \alpha_{uv} \left( \dfrac{u'^+}{2} - v'^+ \right) < -1 \right) \\ 1 & \left( \alpha_{uv} \left( \dfrac{u'^+}{2} - v'^+ \right) > 1 \right) \end{cases} \tag{5.3}$$

where $\alpha_{uv}$ is the rate of the change from wall blowing to suction. As summarized in table 4, the contours representing the control policies have the same inclination angle,

| Case | Control policy | $\alpha_{uv}$ | Drag reduction rate |
|---|---|---|---|
| UV1 | 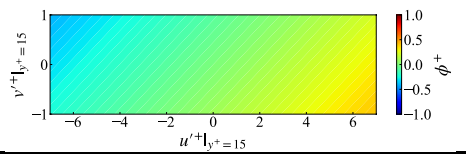 | 0.089 | 17% |
| UV2 | 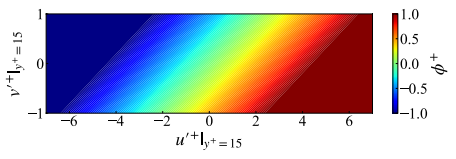 | 0.447 | 27% |
| UV3 | 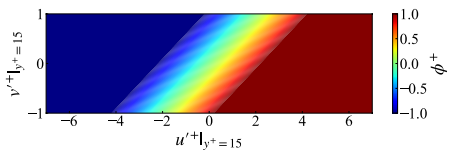 | 0.894 | 35% |
| UV4 | 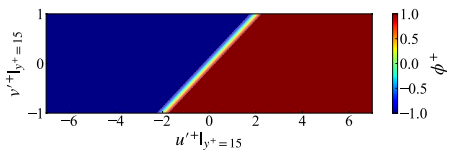 | 8.994 | 35% |

**Table 4:** Control policies and resulting drag reduction rates in the full channel for different slopes $\alpha_{uv}$ from wall blowing to suction in $u'v'$-based control

which is taken from the policy obtained by the reinforcement learning in Case R18 shown in figure 9 (b). It is found that the resultant drag reduction rate increases with increasing $\alpha_{uv}$. This trend is similar to that of $u'$-based control, while opposite to $v'$-based control. From these results, we could conclude that the rapid switch between wall blowing and suction is effective for a policy depending on $u'$ and such a policy can outperform the existing opposition control, which is based on $v'$ only.

5.1.4 Effects of the inclination of the boundary between wall blowing and suction

Finally, we investigate the effects of the inclination angle of the boundary between wall blowing and suction. Specifically, the policies considered here can be expressed by

$$\phi^+ = \begin{cases} -1 \left(0 > \epsilon u'^+|_{y_d^+=15} - v'^+|_{y_d^+=15}\right) \\ 1 \left(0 \le \epsilon u'^+|_{y_d^+=15} - v'^+|_{y_d^+=15}\right) \end{cases} \tag{5.4}$$

where $\epsilon$ controls the inclination angle of the boundary and is changed systematically as shown in table 5.

It is interesting to note that the resultant drag reduction rate increases with increasing ε, i.e., the inclination angle. In Case IA5, where the control policy depends only on $u'$, i.e., $\epsilon = \infty$, the drag reduction rate becomes maximum. This policy is quite similar to those obtained in Cases S18 and LR18 shown in figures 9 (c) and (d), respectively. It can also be seen that the drag reduction rates almost saturate when ε is larger than 0.25. Therefore, the control policies obtained in Cases (b) and (e) can also be considered nearly optimal. Considering that all the policies shown in figure 9 (b-e) are obtained through training in the minimal channel, we can conclude that the reinforcement learning can successfully find the effective control policies which can be transferable to the full channel.

| Case | Control policy | $\epsilon$ | Drag reduction rate |
|---|---|---|---|
| IA1 | 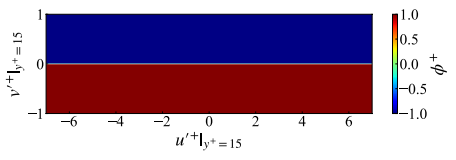 | 0 | 9% |
| IA2 | 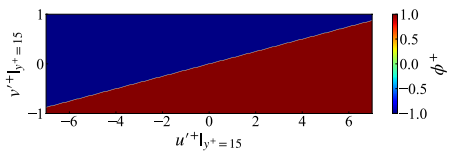 | 0.125 | 17% |
| IA3 | 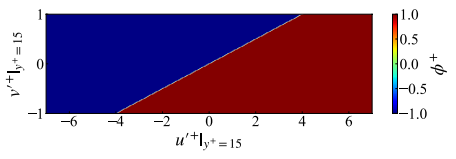 | 0.25 | 27% |
| IA4 | 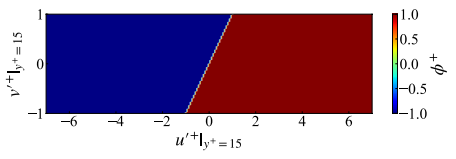 | 1 | 34% |
| IA5 | 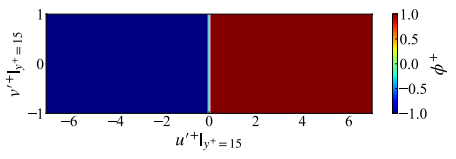 | $\infty$ | 37% |

**Table 5:** Drag reduction rate with different inclination angle of the boundary between rapidly changing wall blowing and suction.

5.2 Spatio-tenporal distribution of control inputs

It is of interest to investigate the spatio-temporal distribution of wall blowing and suction determined by the policy obtained from the current reinforcement learning and how it results in a higher drag reduction rate than that obtained by the conventional opposition control. The instantaneous flow fields as well as the control input at $t^+ = 0.6$ and $t^+ = 20.4$ after the onset of the control in Case R18 are shown in figures 14 (a) and (b), respectively. It can be seen that the control input rapidly switches from wall blowing to suction, i.e., $\phi = 1.0$ and $-1.0$, consistent with the policy shown in figure 9 (b). Just after the onset of the control, at $t^+ = 0.6$, the control input is elongated in the streamwise direction reflecting instantaneous near-wall streaky structures (see, figure 14 (a)). Interestingly, with time passes, the control input transits to a coherent wave-like input as shown in figure 14 (b), which is almost uniform in the spanwise direction and its streamwise wavelength is equal to the streamwise domain size.

We also note that similar wave-like control inputs can be generated when policies with a rapid change from wall blowing and suction are applied. In order to extract a coherent component from the control input, we define the spanwise average of the instantaneous control input $\phi$ on each wall as follows

$$\tilde{\phi}(x,t) = \frac{1}{L_z}\int_0^{L_z} \phi(x,z,t)\, dz \tag{5.5}$$

The spanwise-averaged control inputs in Cases R18, U6 and V4 as functions of $t$ and $x$ are shown in figures 15 (a), (b) and (c), respectively. We note that the corresponding drag reduction rates for Case R18, U6, and V4 are 31%, 37% and 9%, respectively.

In Case R18, the wall blowing and suction switches to the other at a high frequency, while its wave nodes slowly move to upstream (see, figure 15 (a)). In contrast, when the control policy of Case U6 is applied, a downstream traveling wave can be confirmed as shown in figure 15 (b), while a standing wave-like control input can be confirmed in Case V4 (see, figure 15 (c)). Since all the three policies rapidly switch from strong wall blowing to suction depending on the state $u'$ and $v'$ at the detection plane of $y_d^+ = 15$,

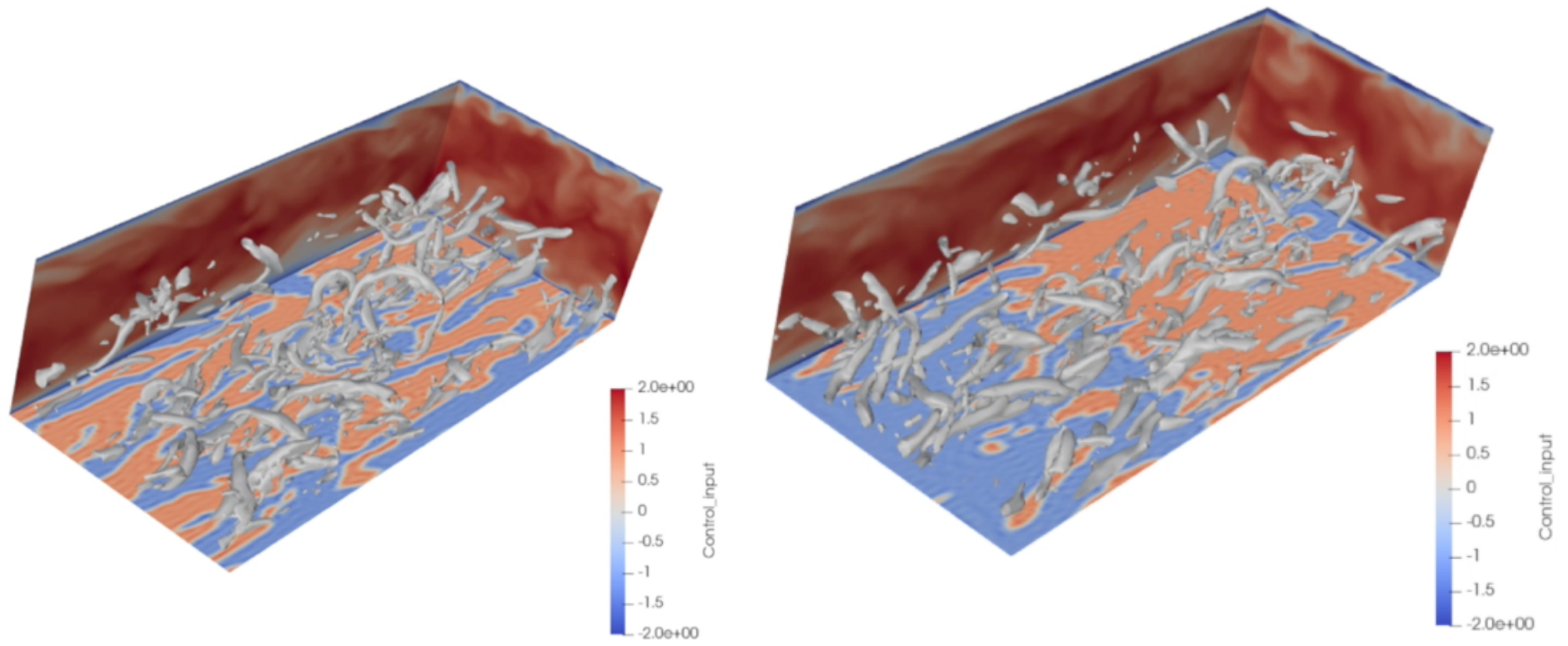

(a) (b)

**Figure 14:** Visualization of the flow fields and the control inputs in the full-size channel when applying the policy obtained in Case R18 (a) at $t^+ = 0.6$ and (b) at $t^+ = 20.4$. White contours show iso-surfaces of the second invariant of the deformation tensor $Q(Q^+ = 0.004)$. Red to blue colors on the bottom wall indicate wall blowing and suction, respectively

such an abrupt change of the control input causes a strong perturbation at the detection plane. This in turn determines the control input in the next time step. Such a feedback between the control input and the flow state at the detection plane should yield the wave-like coherent control inputs observed here. Indeed, the time period of switching from wall blowing to suction in Cases R18 and V4 is equal to the time step for updating the control input, i.e., $\Delta t_{update}^+ = 0.6$.

It has been reported that drag reduction can be achieved by applying a traveling wave-like control input. For example, Min *et al.* (2006) showed that sub-laminar drag can be achieved by an upstream traveling wave of wall blowing and suction when its phase speed is about three times the bulk mean velocity. As mentioned before, the wave nodes obtained in Case R18 shown in figure 15 (a) travels in the upstream direction, and its velocity normalized by the bulk mean velocity is approximately 3.34. Although the traveling speed of the wave node in Case R18 agrees well with the value previously reported, the spanwise-averaged control input obtained in Case R18 switches wall blowing and suction at a higher frequency, so it is essentially different from the upstream traveling waves considered in the previous study.

Meanwhile, relaminarization is also caused by a downstream traveling wave when the phase speed is larger than 1.5 times the bulk mean velocity (Lieu *et al.* 2010; Mamori *et al.* 2014). However, the phase-speed of the downstream wave obtained in Case U6 shown figure 15 (b) is much faster, i.e., around 28 when it is normalized by the bulk mean velocity. In order to clarify whether the coherent wave-like inputs observed in figure 15 (a-c) alone lead to drag reduction or not, we conduct additional simulations where only the coherent control inputs, which are uniform in the spanwise direction and only depend on the streamwise direction and the time, are applied. It should be noted that, in these cases, the applied controls are no longer feedback controls, but predetermined controls, since the control inputs are determined a priori, and do not depend on the instantaneous flow state. The results are shown in figure 16. It is found that solely applying the coherent wave-like control inputs hardly leads to drag reduction. This indicates that the present controls are feedback controls and applying a control input based on the instantaneous flow state is essential to yield the drag reduction effects.

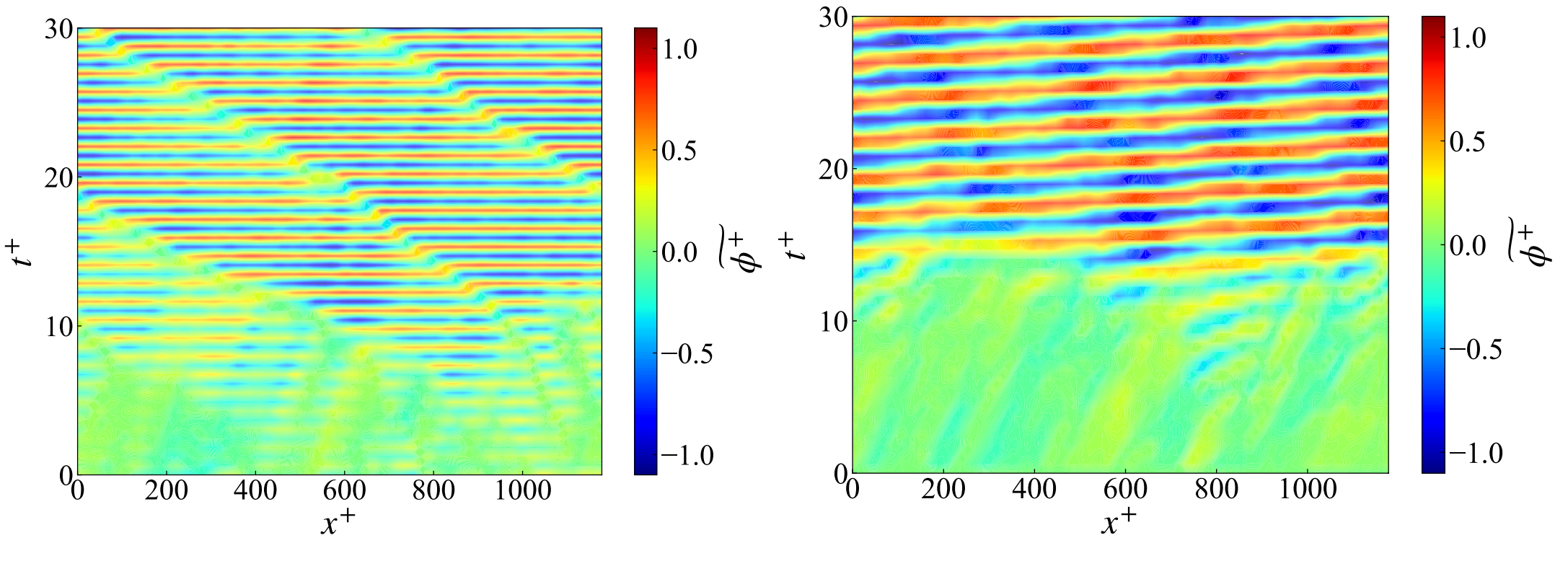

(a) (b)

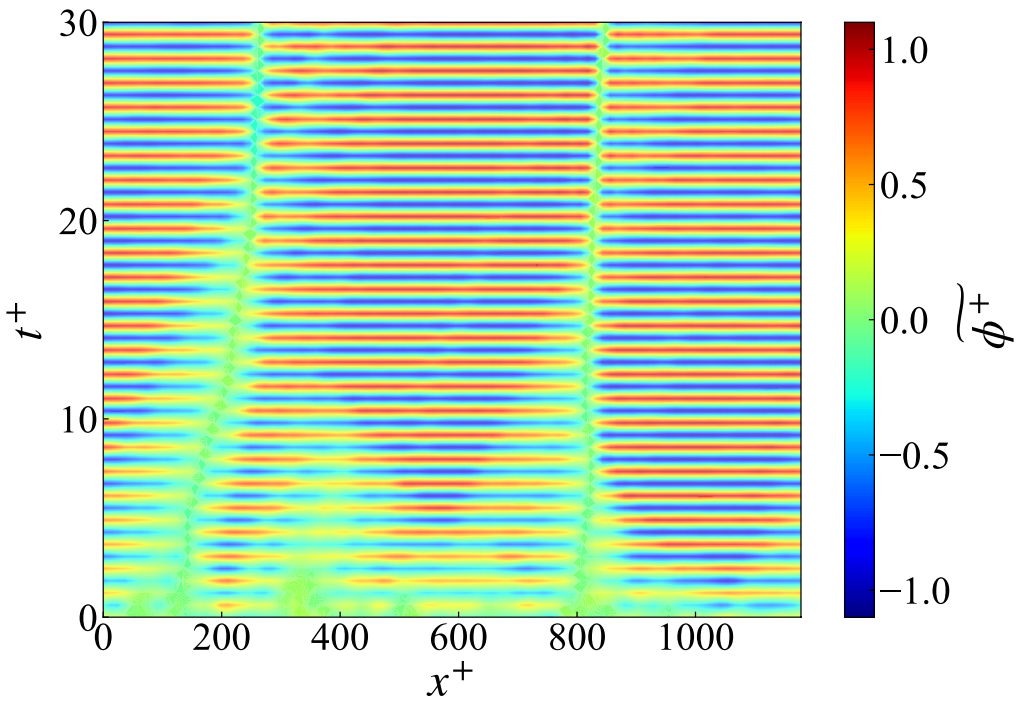

(c)

**Figure 15:** Spanwise-averaged control input $\widetilde{\phi^+}$ as a function of time t and the streamwise coordinate x in (a) Case R18, (b) Case U6, (c) Case V4.

## 6 Conclusions

In this study, reinforcement learning is first applied to obtain effective control strategies using wall blowing and suction for reducing skin frictional drag in a fully developed turbulent channel flow. The present framework is based on the Deep Deterministic Policy Gradient (DDPG) algorithm (Lillicrap *et al.* 2015), where the actor network dictating a control policy reads the flow state and outputs the action, i.e., the control input, while the critic network estimates the expected total future reward, i.e., a long-term drag reduction rate, when a certain action is taken under a certain flow state. The two networks are simultaneously trained through a number of trials in direct numerical simulation.

We first considered a simple policy where the local wall blowing and suction is linearly related to the wall-normal velocity fluctuation at the detection plane of $y_d^+ = 15$. It is found that the current reinforcement learning successfully finds the optimal weight coefficient reported in the previous study (Chung & Talha 2011). Next, we extended the above framework by adding the streamwise velocity fluctuation as well as the wall-normal velocity fluctuation as the state, and also including non-linear activation functions in the actor network. It is demonstrated that the obtained policies lead to drag reduction rates as high as 37%, which is higher than 23 % achieved by the existing opposition control. The obtained control policies are characterized by a sharp change from wall blowing to suction depending on the streamwise and wall-normal velocity fluctuations at the detection plane. Further detailed analyses indicate that such a control policy with a rapid switch between wall blowing and suction is particularly effective when a control policy depends on the streamwise velocity fluctuation at the detection plane.

It should be emphasized that finding such an effective and highly non-linear control policy is quite difficult by solely relying on researchers' insights, and it becomes pos- sible by a systematic learning framework leveraged by neural networks. One of great advantages in the reinforcement learning is that it can learn not only from successes, but also from failures through numerous trials. In the flow control community, effective control laws have often been sought by human through trials and errors. Reinforcement learning has a potential to replace such human efforts to explore effective control policies Although we are still in the process of developing newly emerging methodologies, based on the obtained control policies, it is expected that we will be able to gain a deeper understanding of flow physics and new control guidelines. The unique control policies obtained in the present study would also contribute to these purposes. In the current study, we only consider the streamwise and wall-normal velocity fluctuations at a certain distance from the wall as a state, and there is a possibility that more effective control strategies could be found by extending the state in space and/or time. Meanwhile, our preliminary results suggest that the learning becomes more difficult when the network size becomes larger (see, Appendix B). Establishing effective learning methodologies is obviously crucial. Also, the current study only considers a single low Reynolds number, and the applicability of the current approach to higher Reynolds numbers needs to be investigated. They should be tackled in future studies.

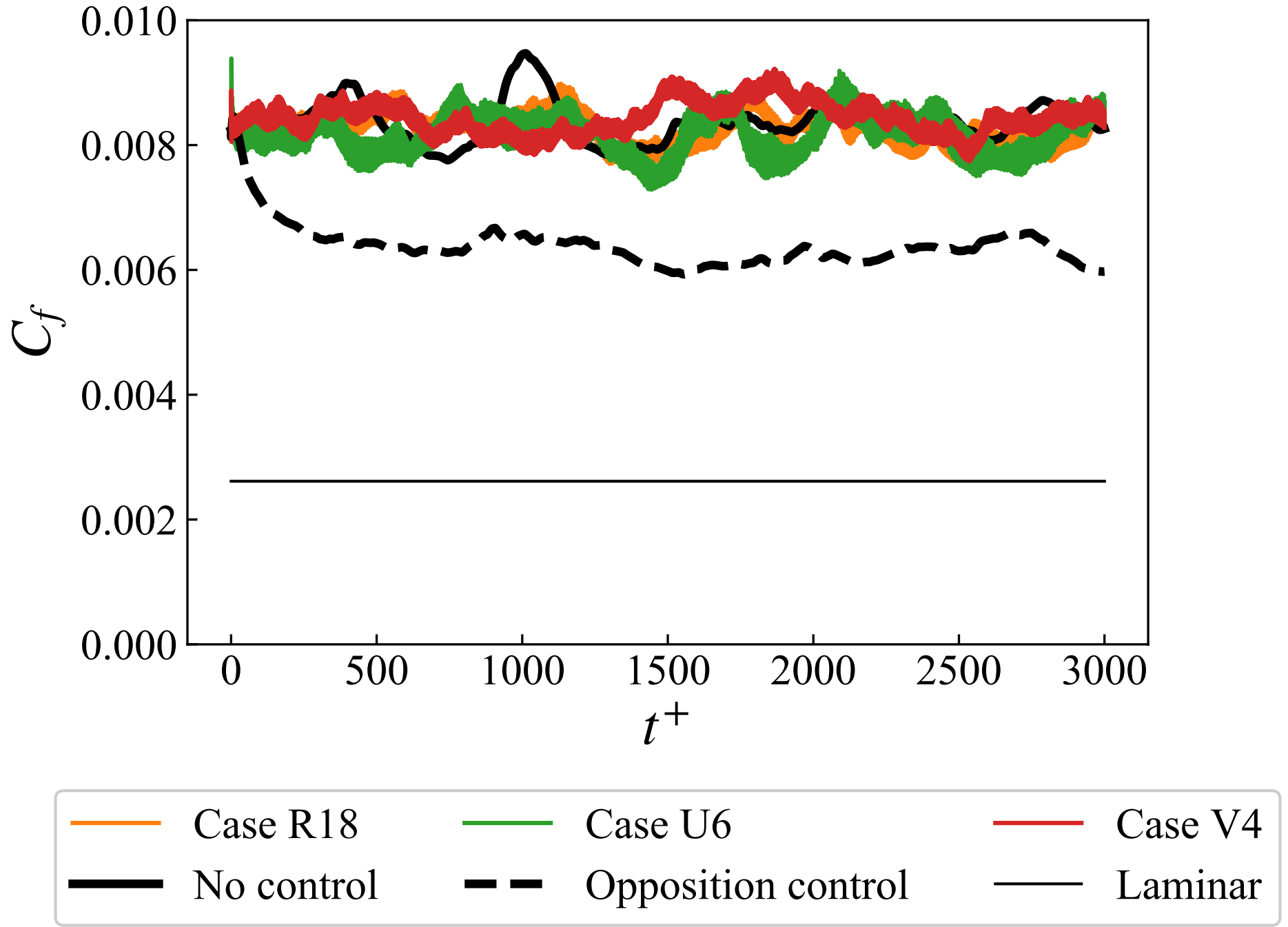

**Figure 16:** Time evolutions of $C_f$ obtained with the spanwise-averaged control inputs in Cases R18, U6 and V4. For comparison, the values in the uncontrolled flow, the opposition control, and the laminar flow are also plotted as thick black, broken thick black and thin black lines, respectively.

## Appendix A: Control policy based on flow states at different locations

The present results indicate that control policies based on the steamwise velocity fluctuation are generally more effective than those based on the wall-normal velocity fluctuation only. Meanwhile, all the policies presented in the present study are based on the flow state at a single location of $y_d^+ = 15$from the wall. Here, we show one example where the state is extended to multiple locations from the wall.

Specifically, we use the streamwise velocity fluctuation $u'$ at 10 different locations from the wall, i.e., $y_{d,i}^+ = 5,10,15,20,25,30,34,41$ and 51. In order to make a problem simple, we consider the following linear control policy which can be expressed as

$$a \equiv \phi(x,z,t) = \tanh\left\{\sum_{i=1}^{10} \alpha_i u'(x, y_{d,i}, z) + \right\} + N \qquad (A1)$$

Where $\alpha_i (i = 1, \cdots 10)$ and β are linear weights and a bias to be optimized, while N is a zero-mean random noise, the standard deviation of which is 0.1 in the wall unit.

Figure 17 shows the distribution of the weights $\alpha_i$ for different distances from the wall at the end of the episode where the maximum drag reduction is achieved. It can be seen that the weights around $y_d^+ = 15 \sim 20$ become maximum. When this policy is applied in the full channel, the drag reduction rate of 36% is achieved. This value is similar to those obtained in the present study where a single detection plane at $y_d^+ = 15$ is considered. Therefore, it is expected that increasing the number of detection planes will not significantly further improve the control performance.

## Appendix B: Effects of the numbers of layers and nodes employed in the actor

Here, we summarize some results with different numbers of layers and nodes used for the actor. The obtained control policies and resulting drag reduction rates for all the cases considered are summarized in table 6. Except for the numbers of layers and nodes in the actor, the other settings such as an activation function in the actor, the hyper parameters in the critic and learning procedures are the same as Case R18 in table 1. The drag reduction rates listed in table 6 are obtained by averaging the final 20 episodes during the training after the flow fields converge to equilibrium states.

It can be seen that the obtained control policies are qualitatively similar in Cases R14, R18 and R24. Among them, Case R18 results in the highest drag reduction rate. In Case R28, where the actor has the most complex network among all the cases considered, drag reduction is not achieved. Hence, we conclude that Case R18 with 1 layer and 8 nodes is suitable for the present problem setting.

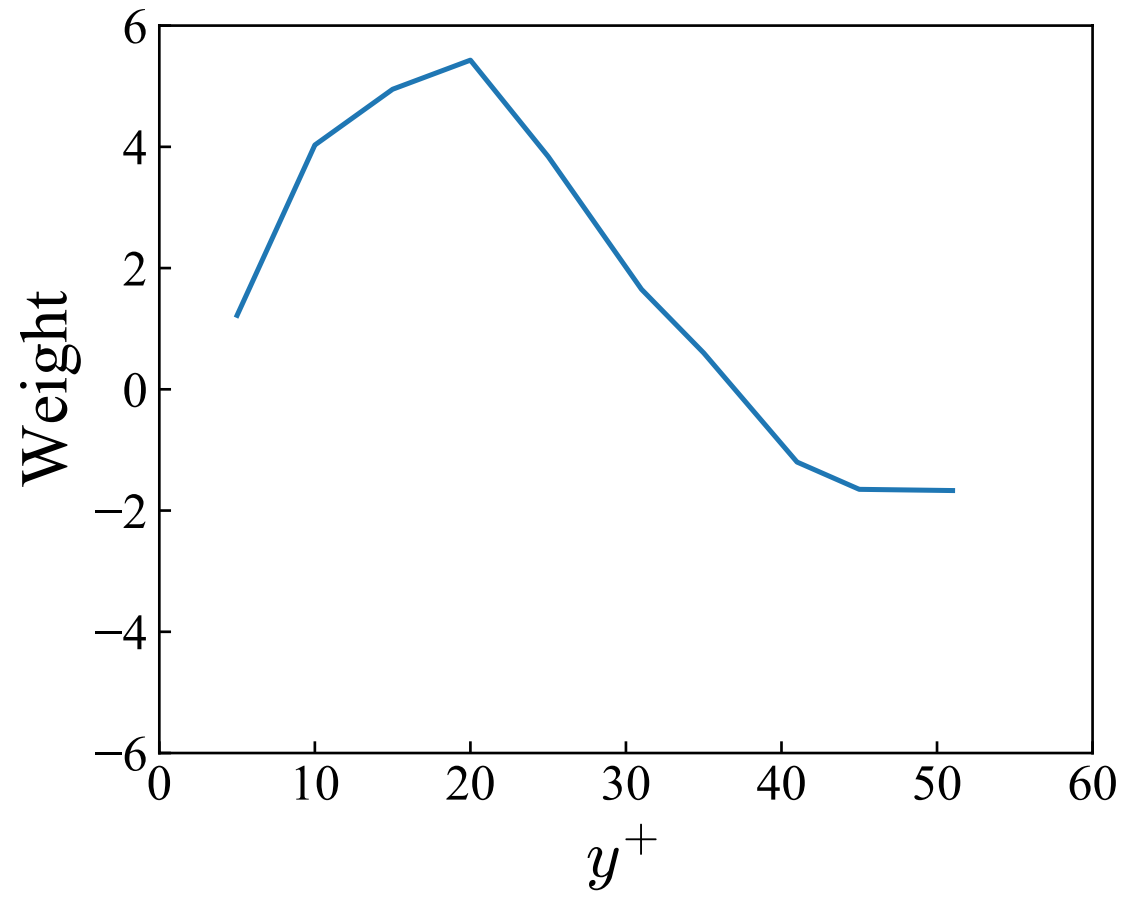

**Figure 17:** Weights for $u'$ at different locations in the best policy obtained for equation (A 1)

| Case | Layers | Nodes | policy | Drag reduction rate |
|---|---|---|---|---|
| R14 | 1 | 4 | 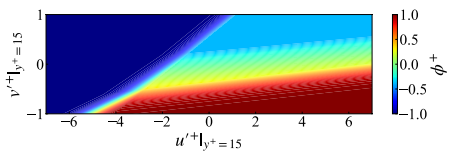 | 9% |
| R18 | 1 | 8 | 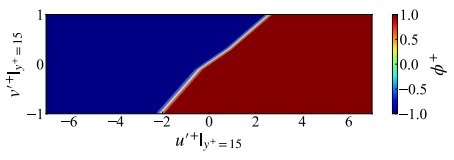 | 17% |
| R24 | 2 | 4 | 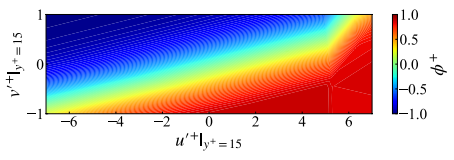 | 27% |
| R28 | 2 | 8 | 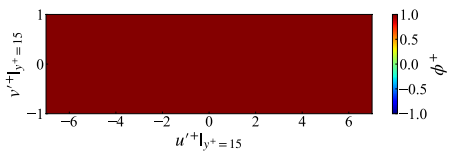 | 34% |

**Table 6:** Cases with different numbers of layers and nodes used in the hidden layers of the actor

## Appendix C: Dependency of control performance on numerical schemes

As shown in figure 9, effective policies obtained in the present study commonly show a rapid change from wall blowing to suction depending on the flow state at $y_d^+ = 15$. This may cause unphysical oscillations and affect the resulting drag reduction rate. Especially, such numerical effects could appear more strongly in a spectral method employed in the present study due to the Gibbs phenomena. Therefore, we conduct additional simulations with a finite difference method in order to confirm the universality of the present results.

Specifically, we use an open-source flow solver called Incompact3d (Laizet & Lamballais 2009; Laizet & Li 2011), which is based on the sixth-order compact finite difference scheme. Time integration is conducted by using the second-order Crank-Nicolson scheme for the wall-normal diffusion term, whereas the third-order Adams-Bashforth scheme are applied for the other terms. The friction Reynolds number and the domain size are set to $Re_\tau \approx 150$and$(L_x, L_y, L_z) = (2.5\pi, 2.0, \pi)$., respectively. These are the same as those used for the full channel in the present study. The number of grids in each direction is set to$(N_x, N_y, N_z) = (128,129,96)$, resulting in the spatial resolutions of $\Delta x^+ = 9.2$, $\Delta y^+ = 0.83 \sim 6.6$,and $\Delta z^+ = 4.9$.

Table 7 shows the comparisons of the drag reduction rates obtained by the present pseudo-spectral and finite difference methods for typical control policies, i.e., Cases R18, U3, U6, and V4, together with the results of the opposition control. We note that rapid switch from wall blowing to suction exists in Cases R18, U6 and V4, while it changes smoothly in the rest of the cases. It can be seen that the impacts of the employed numerical schemes on the resultant drag reduction rates generally remain minor, so that the control performances obtained by the preset policies can be considered universal.

| Case | Control policy | Pseudo-spectral | Finite difference |
|---|---|---|---|
| Opposition contol | 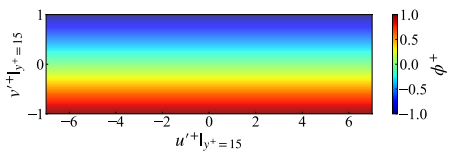 | 23% | 25% |
| R18 | 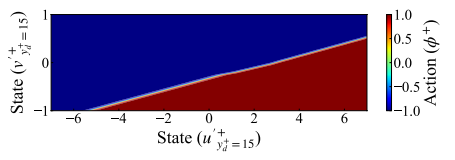 | 31% | 35% |
| U3 | 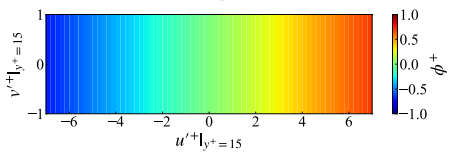 | 20% | 21% |
| U6 | 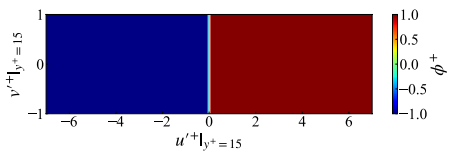 | 37% | 37% |
| V4 | 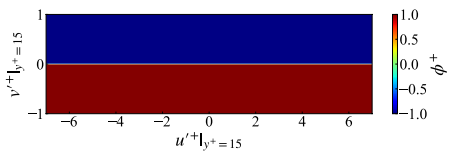 | 9% | 12% |
| UV2 | 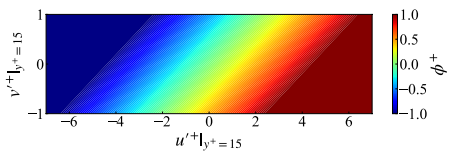 | 27% | 23% |

**Table 7:** Comparison of drag reduction rates for different policies obtained by pseudo-spectral and finite difference methods.

## References

Beintema, G., Corbetta, A., Biferale, L. & Toschi, F. 2020 Controlling rayleigh–benard convection via reinforcement learning. J. Turbul. 21 (9–10), 585–605.

Bewley, T., Moin, P. & Temam, R. 2001 Dns-based predictive control of turbulence: an optimal benchmark for feedback algorithms. J. Fluid Mech. 447 (2), 179–225.

Boyd, J. P. 2001 Chebyshev and Fourier spectral methods. Courier Corporation.

Brunton, S. L. & Noack, B. R. 2015 Closed-loop turbulence control: Progress and challenges. Appl. Mech. Rev. 67 (5).

Choi, H., Moin, P. & Kim, J. 1994 Active turbulence control for drag reduction in wall-bounded flows. J. Fluid Mech. 262, 75–110.

Choi, H., Temam, R., Moin, P. & Kim, J. 1993 Feedback control for unsteady flow and its application to the stochastic burgers equation. J. Fluid Mech. 253, 509–543.

Chung, Y. M. & Talha, T. 2011 Effectiveness of active flow control for turbulent skin friction drag reduction. Phys. Fluids 23 (2), 025102.

Dean, B. & Bhushan, B. 2010 Shark-skin surfaces for fluid-drag reduction in turbulent flow: a review. Philos. Trans. Royal Soc. A 368 (1929), 4775–4806.

Dressler, O. J., Howes, P. D., Choo, J. & deMello, A. J. 2018 Reinforcement learning for dynamic microfluidic control. ACS Omega 3 (8), 10084–10091.

Fan, D., Yang, L., Wang, Z., Triantafyllou, M. S. & Karniadakis, G. E. 2020 Reinforcement learning for bluff body active flow control in experiments and simulations. Proc. Natl. Acad. Sci. U.S.A. 117 (42), 26091–26098.

Fukagata, K., Iwamoto, K. & Kasagi, N. 2002 Contribution of reynolds stress distribution to the skin friction in wall-bounded flows. Phys. Fluids 14 (11), L73–L76.

Ghraieb, H., Viquerat, J., Larcher, A., Meliga, P. & Hachem, E. 2021 Single-step deep reinforcement learning for open-loop control of laminar and turbulent flows. Phys. Rev. Fluids 6 (5), 053902.

Hachem, E., Ghraieb, H., Viquerat, J., Larcher, A. & Meliga, P. 2021 Deep reinforcement learning for the control of conjugate heat transfer. J. Comput. Phys. 436, 110317.

Gad-el Hak, M. 1996 Modern developments in flow control. Appl. Mech. Rev. 49, 365–379.

Hammond, E. P., Bewley, T. R. & Moin, P. 1998 Observed mechanisms for turbulence attenuation and enhancement in opposition-controlled wall-bounded flows. Phys. Fluids 10 (9), 2421–2423.

Hasegawa, Y. & Kasagi, N. 2011 Dissimilar control of momentum and heat transfer in a fully developed turbulent channel flow. J. Fluid Mech. 683, 57–93.

Jim´enez, J. & Moin, P. 1991 The minimal flow unit in near-wall turbulence. J. Fluid Mech. 225, 213–240.

Jung, W. J., Mangiavacchi, N. & Akhavan, R. 1992 Suppression of turbulence in wall-bounded flows by high-frequency spanwise oscillations. Phys. Fluids A 4 (8), 1605–1607.

Kametani, Y. & Fukagata, K. 2011 Direct numerical simulation of spatially developing turbulent boundary layers with uniform blowing or suction. J. Fluid Mech. 681, 154–172.
Kim, J. & Moin, P. 1985 Application of a fractional-step method to incompressible Navier-Stokes equations. J. Comp. Phys. 59 (2), 308–323.
Kim, K. C. & Adrian, R. J. 1999 Very large-scale motion in the outer layer. Phys. Fluids 11 (2), 417–422.
Kober, J., Bagnell, J. A. & Peters, J. 2013 Reinforcement learning in robotics: A survey. Int. J. Robot. Res. 32 (11), 1238–1274.
Koizumi, H., Tsutsumi, S. & Shima, E. 2018 Feedback control of karman vortex shedding from a cylinder using deep reinforcement learning. 2018 Flow Control Conference p. 3691.
Laizet, S. & Lamballais, E. 2009 High-order compact schemes for incompressible flows: A simple and efficient method with quasi-spectral accuracy. J. Comp. Phys. 228 (16), 5989–6015.
Laizet, S. & Li, N. 2011 Incompact3d: A powerful tool to tackle turbulence problems with up to O(105) computational cores. Intl. J. Numer. Methods Fluids 67 (11), 1735–1757.
Lee, C., Kim, J. & Choi, H. 1998 Suboptimal control of turbulent channel flow for drag reduction. J. Fluid Mech. 358, 245–258.
Lee, X. Y., Balu, A., Stoecklein, D., Ganapathysubramanian, B. & Sarkar, S. 2021 A case study of deep reinforcement learning for engineering design: Application to microfluidic devices for flow sculpting. J. Mech. Des. 141 (11), 111401.
Li, R., Zhang, Y. & Chen, H. 2021 Learning the aerodynamic design of supercritical airfoils through deep reinforcement learning. AIAA J. 59 (10), 3988–4001.
Lieu, B. K., Marref, R. & Jovanovi´c, M. R. 2010 Controlling the onset of turbulence by streamwise travelling waves. part 2. direct numerical simulation. J. Fluid Mech. 663, 100–119.
Lillicrap, T. P., Hunt, J. J., Pritzel, A., Heess, N., Erez, T., Tassa, Y., Silver, D. & Wierstra, D. 2015 Continuous control with deep reinforcement learning. arXiv preprint arXiv:1509.02971.
Mamori, H., Iwamoto, K. & Murata, A. 2014 Effect of the parameters of traveling waves created by blowing and suction on the relaminarization phenomena in fully developed turbulent channel flow. Phys. Fluids 26 (1), 015101.
Min, T., Kang, S. M., Speyer, J. L. & Kim, J. 2006 Sustained sub-laminar drag in a fully developed channel flow. J. Fluid Mech. 558, 309–318.
Novati, G., Verma, S., Alexeev, D., Rossinelli, D., Van Rees, W. M. & Koumoutsakos, P. 2018 Synchronisation through learning for two self-propelled swimmers. Bioinspir Biomim 12 (3), 036001.
Paris, R., Beneddine, S. & Dandois, J. 2021 Robust flow control and optimal sensor placement using deep reinforcement learning. J. Fluid Mech. 913.

Qin, S., Wang, S., Wang, L., Wang, C., Sun, G. & Zhong, Y. 2021 Multi-objective optimization of cascade blade profile based on reinforcement learning. Appl. Sci. 11 (1), 106.

Quadrio, M. & Ricco, P. 2004 Critical assessment of turbulent drag reduction through spanwise wall oscillations. J. Fluid Mech. 521, 251.

Rabault, J., Kuchta, M., Jensen, A., R´eglade, U. & Cerardi, N. 2019 Artificial neural networks trained through deep reinforcement learning discover control strategies for active flow control. J. Fluid Mech. 865, 281–302.

Rabault, J. & Kuhnle, A. 2019 Accelerating deep reinforcement learning strategies of flow control through a multi-environment approach. Phys. Fluids 31 (9), 094105.

Ren, F., Rabault, J. & Tang, H. 2021 Applying deep reinforcement learning to active flow control in weakly turbulent conditions. Phys. Fluids 33 (3), 037121.

Silver, D., Huang, A., Maddison, C. J., Guez, A., Sifre, L., Van Den Driessche, G., Schrittwieser, D., Antonoglou, I., Panneershelvam, V., Lanctot, M., Dieleman, S., Grewe, D., Nham, J., Kalchbrenner, N., Sutskever, I., Lillicrap, T., Leach, M., Kavukcuoglu, K., Graepel, T. & Hassabis, D. 2016 Mastering the game of go with deep neural networks and tree search. Nature 529 (7587), 484–489.

Sumitani, Y. & Kasagi, N. 1995 Direct numerical simulation of turbulent transport with uniform wall injection and suction. AIAA J. 33 (7), 1220–1228.

Sutton, R. S. & Barto, A. G. 2018 Reinforcement learning: An introduction. MIP press.

Suzuki, T. & Hasegawa, Y. 2017 Estimation of turbulent channel flow at based on the wall measurement using a simple sequential approach. J. Fluid Mech. 830, 760–796.

Tang, H., Rabault, J., Kuhnle, A., Wang, Y. & Wang, T. 2020 Robust active flow control over a range of reynolds numbers using an artificial neural network trained through deep reinforcement learning. Phys. Fluids 32 (5), 053605.

Tokarev, M., Palkin, E. & Mullyadzhanov, R. 2020 Deep reinforcement learning control of cylinder flow using rotary oscillations at low reynolds number. Energies 13 (22), 5920.

Verma, S., Novati, G. & Koumoutsakos, P. 2018 Efficient collective swimming by harnessing vortices through deep reinforcement learning. Proc. Natl. Acad. Sci. U.S.A. 115 (23), 5849–5854.

Viquerat, J., Rabault, J., Kuhnle, A., Ghraieb, H., Larcher, A. & Hachem, E. 2021 Direct shape optimization through deep reinforcement learning. J. Comput. Phys. 428, 110080.

Wang, Q., Hu, R. & Blonigan, P. 2014 Least squares shadowing sensitivity analysis of chaotic limit cycle oscillations. J. Comp. Phys. 267, 210–224.

Xu, H., Zhang, W., Deng, J. & Rabault, J. 2020 Active flow control with rotating cylinders by an artificial neural network trained by deep reinforcement learning. J. Hydrodyn. 32 (2), 254–258.

Yamamoto, A., Hasegawa, Y. & Kasagi, N. 2013 Optimal control of dissimilar heat and momentum transfer in a fully developed turbulent channel flow. J. Fluid Mech 733, 189–220.

Yan, L., Chang, X., Tian, R., Wang, N., Zhang, L. & Liu, W. 2020 A numerical simulation method for bionic fish self-propelled swimming under control based on deep reinforcement learning. Proc. Inst. Mech. Eng. C., J. Mech. Eng. Sci. 234 (17), 3397–3415.

Yan, X., Zhu, J., Kuang, M. & Wang, X. 2019 Aerodynamic shape optimization using a novel optimizer based on machine learning techniques. Aerosp. Sci. Technol. 86, 826–835.

Zhu, Y., Tian, F. B., Young, J., Liao, J. C. & Lai, J. 2021 A numerical study of fish adaption behaviors in complex environments with a deep reinforcement learning and immersed boundary–lattice boltzmann method. Sci. Rep. 11 (1), 1–20.